\documentclass[twocolumn,superscriptaddress,showpacs]{revtex4-1}

\usepackage{graphicx}	
\usepackage{amsmath}	
\usepackage{amssymb}	
\usepackage{bm}		
\usepackage{color}
\usepackage{exscale}
\usepackage{float}
\usepackage{natbib}
\usepackage{relsize}
\usepackage{tabularx}
\usepackage{longtable}
\usepackage{ltxtable}
\usepackage{bbm}
\usepackage{verbatim}
\usepackage{enumerate}
\usepackage{enumitem}

\def\apj{ApJ}%
\def\aap{A\&A}%
\def\mnras{MNRAS}%
\def\prd{Phys.~Rev.~D}%
\def\pasp{PASP}%
\def\ssr{Space~Sci.~Rev.}%
\def\jcap{JCAP}%
\def\physrep{Phys.~Rep.}%
\newcommand{\tbf}{\textbf}

\newcommand{\bea}{\begin{eqnarray}}
\newcommand{\be}{\begin{equation}}
\newcommand{\ben}{\begin{enumerate}}
\newcommand{\bi}{\begin{itemize}}
\newcommand{\eea}{\end{eqnarray}}
\newcommand{\ee}{\end{equation}}
\newcommand{\ei}{\end{itemize}}
\newcommand{\een}{\end{enumerate}}

\newcommand{\matC}{\mathbf C}

\newcommand{\like}{L}

\newcommand{\p}{\vek p}

\newcommand{\D}{\vek D}

\newcommand{\M}{\vek M}

\newcommand{\farcm}{\mbox{\ensuremath{.\mkern-4mu^\prime}}}
\newcommand{\om}{\Omega_\mr m}
\newcommand{\omb}{\Omega_\mr b}

\newcommand{\ns}{n_s}

\newcommand{\mr}{\mathrm}

\newcommand{\vek}{\mathbf}

\newcommand{\emcee}{{\sc Emcee}}

\begin{document}

\title{Dark Energy Survey Year 1 Results: Multi-Probe Methodology and Simulated Likelihood Analyses}
\author{E.~Krause}\email[Corresponding author: ]{lise@slac.stanford.edu}
\affiliation{Kavli Institute for Particle Astrophysics \& Cosmology, P. O. Box 2450, Stanford University, Stanford, CA 94305, USA}
\author{T.~F.~Eifler}\email[Corresponding author: ]{tim.eifler@jpl.nasa.gov}
\affiliation{Department of Physics, California Institute of Technology, Pasadena, CA 91125, USA}
\affiliation{Jet Propulsion Laboratory, California Institute of Technology, 4800 Oak Grove Dr., Pasadena, CA 91109, USA}
\author{J.~Zuntz}
\affiliation{Scottish Universities Physics Alliance, Institute for Astronomy, University of Edinburgh, Edinburgh EH9 3HJ, UK}
\author{O.~Friedrich}
\affiliation{Max Planck Institute for Extraterrestrial Physics, Giessenbachstrasse, 85748 Garching, Germany}
\affiliation{Universit\"ats-Sternwarte, Fakult\"at f\"ur Physik, Ludwig-Maximilians Universit\"at M\"unchen, Scheinerstr. 1, 81679 M\"unchen, Germany}
\author{M.~A.~Troxel}
\affiliation{Center for Cosmology and Astro-Particle Physics, The Ohio State University, Columbus, OH 43210, USA}
\affiliation{Department of Physics, The Ohio State University, Columbus, OH 43210, USA}
\author{S.~Dodelson}
\affiliation{Fermi National Accelerator Laboratory, P. O. Box 500, Batavia, IL 60510, USA}
\affiliation{Kavli Institute for Cosmological Physics, University of Chicago, Chicago, IL 60637, USA}
\author{J.~Blazek}
\affiliation{Center for Cosmology and Astro-Particle Physics, The Ohio State University, Columbus, OH 43210, USA}
\affiliation{Laboratory of Astrophysics, \'Ecole Polytechnique F\'ed\'erale de Lausanne (EPFL), Observatoire de Sauverny, 1290 Versoix, Switzerland}
\author{L.~F.~Secco}
\affiliation{Department of Physics and Astronomy, University of Pennsylvania, Philadelphia, PA 19104, USA}
\author{N.~MacCrann}
\affiliation{Center for Cosmology and Astro-Particle Physics, The Ohio State University, Columbus, OH 43210, USA}
\affiliation{Department of Physics, The Ohio State University, Columbus, OH 43210, USA}
\author{E.~Baxter}
\affiliation{Department of Physics and Astronomy, University of Pennsylvania, Philadelphia, PA 19104, USA}
\author{C.~Chang}
\affiliation{Kavli Institute for Cosmological Physics, University of Chicago, Chicago, IL 60637, USA}
\author{N.~Chen}
\affiliation{Kavli Institute for Cosmological Physics, University of Chicago, Chicago, IL 60637, USA}
\author{M.~Crocce}
\affiliation{Institut de Ci\`encies de l'Espai, IEEC-CSIC, Campus UAB, Carrer de Can Magrans, s/n,  08193 Bellaterra, Barcelona, Spain}
\author{J.~DeRose}
\affiliation{Department of Physics, Stanford University, 382 Via Pueblo Mall, Stanford, CA 94305, USA}
\affiliation{Kavli Institute for Particle Astrophysics \& Cosmology, P. O. Box 2450, Stanford University, Stanford, CA 94305, USA}
\author{A.~Fert\'e}
\affiliation{Institute for Astronomy, University of Edinburgh, Edinburgh EH9 3HJ, UK}
\author{N.~Kokron}
\affiliation{ICTP South American Institute for Fundamental Research\\ Instituto de F\'{\i}sica Te\'orica, Universidade Estadual Paulista, S\~ao Paulo, Brazil}
\affiliation{Laborat\'orio Interinstitucional de e-Astronomia - LIneA, Rua Gal. Jos\'e Cristino 77, Rio de Janeiro, RJ - 20921-400, Brazil}
\author{F.~Lacasa}
\affiliation{D\'{e}partement de Physique Th\'{e}orique and Center for Astroparticle Physics, Universit\'{e} de Gen\`{e}ve, 24 quai Ernest Ansermet, CH-1211 Geneva, Switzerland}
\affiliation{Laborat\'orio Interinstitucional de e-Astronomia - LIneA, Rua Gal. Jos\'e Cristino 77, Rio de Janeiro, RJ - 20921-400, Brazil}
\author{V. Miranda}
\affiliation{Department of Physics and Astronomy, University of Pennsylvania, Philadelphia, PA 19104, USA}
\author{Y.~Omori}
\affiliation{Department of Physics and McGill Space Institute, McGill University, Montreal, Quebec H3A 2T8, Canada}
\author{A.~Porredon}
\affiliation{Institut de Ci\`encies de l'Espai, IEEC-CSIC, Campus UAB, Carrer de Can Magrans, s/n,  08193 Bellaterra, Barcelona, Spain}
\author{R.~Rosenfeld}
\affiliation{ICTP South American Institute for Fundamental Research\\ Instituto de F\'{\i}sica Te\'orica, Universidade Estadual Paulista, S\~ao Paulo, Brazil}
\affiliation{Laborat\'orio Interinstitucional de e-Astronomia - LIneA, Rua Gal. Jos\'e Cristino 77, Rio de Janeiro, RJ - 20921-400, Brazil}
\author{S.~Samuroff}
\affiliation{Jodrell Bank Center for Astrophysics, School of Physics and Astronomy, University of Manchester, Oxford Road, Manchester, M13 9PL, UK}
\author{M.~Wang}
\affiliation{Fermi National Accelerator Laboratory, P. O. Box 500, Batavia, IL 60510, USA}
\author{R.~H.~Wechsler}
\affiliation{Department of Physics, Stanford University, 382 Via Pueblo Mall, Stanford, CA 94305, USA}
\affiliation{Kavli Institute for Particle Astrophysics \& Cosmology, P. O. Box 2450, Stanford University, Stanford, CA 94305, USA}
\affiliation{SLAC National Accelerator Laboratory, Menlo Park, CA 94025, USA}
\author{T. M. C.~Abbott}
\affiliation{Cerro Tololo Inter-American Observatory, National Optical Astronomy Observatory, Casilla 603, La Serena, Chile}
\author{F.~B.~Abdalla}
\affiliation{Department of Physics \& Astronomy, University College London, Gower Street, London, WC1E 6BT, UK}
\affiliation{Department of Physics and Electronics, Rhodes University, PO Box 94, Grahamstown, 6140, South Africa}
\author{S.~Allam}
\affiliation{Fermi National Accelerator Laboratory, P. O. Box 500, Batavia, IL 60510, USA}
\author{J.~Annis}
\affiliation{Fermi National Accelerator Laboratory, P. O. Box 500, Batavia, IL 60510, USA}
\author{K.~Bechtol}
\affiliation{LSST, 933 North Cherry Avenue, Tucson, AZ 85721, USA}
\author{A.~Benoit-L{\'e}vy}
\affiliation{CNRS, UMR 7095, Institut d'Astrophysique de Paris, F-75014, Paris, France}
\affiliation{Department of Physics \& Astronomy, University College London, Gower Street, London, WC1E 6BT, UK}
\affiliation{Sorbonne Universit\'es, UPMC Univ Paris 06, UMR 7095, Institut d'Astrophysique de Paris, F-75014, Paris, France}
\author{G.~M.~Bernstein}
\affiliation{Department of Physics and Astronomy, University of Pennsylvania, Philadelphia, PA 19104, USA}
\author{D.~Brooks}
\affiliation{Department of Physics \& Astronomy, University College London, Gower Street, London, WC1E 6BT, UK}
\author{D.~L.~Burke}
\affiliation{Kavli Institute for Particle Astrophysics \& Cosmology, P. O. Box 2450, Stanford University, Stanford, CA 94305, USA}
\affiliation{SLAC National Accelerator Laboratory, Menlo Park, CA 94025, USA}
\author{D.~Capozzi}
\affiliation{Institute of Cosmology \& Gravitation, University of Portsmouth, Portsmouth, PO1 3FX, UK}
\author{M.~Carrasco~Kind}
\affiliation{Department of Astronomy, University of Illinois, 1002 W. Green Street, Urbana, IL 61801, USA}
\affiliation{National Center for Supercomputing Applications, 1205 West Clark St., Urbana, IL 61801, USA}
\author{J.~Carretero}
\affiliation{Institut de F\'{\i}sica d'Altes Energies (IFAE), The Barcelona Institute of Science and Technology, Campus UAB, 08193 Bellaterra (Barcelona) Spain}
\author{C.~B.~D'Andrea}
\affiliation{Department of Physics and Astronomy, University of Pennsylvania, Philadelphia, PA 19104, USA}
\author{L.~N.~da Costa}
\affiliation{Laborat\'orio Interinstitucional de e-Astronomia - LIneA, Rua Gal. Jos\'e Cristino 77, Rio de Janeiro, RJ - 20921-400, Brazil}
\affiliation{Observat\'orio Nacional, Rua Gal. Jos\'e Cristino 77, Rio de Janeiro, RJ - 20921-400, Brazil}
\author{C.~Davis}
\affiliation{Kavli Institute for Particle Astrophysics \& Cosmology, P. O. Box 2450, Stanford University, Stanford, CA 94305, USA}
\author{D.~L.~DePoy}
\affiliation{George P. and Cynthia Woods Mitchell Institute for Fundamental Physics and Astronomy, and Department of Physics and Astronomy, Texas A\&M University, College Station, TX 77843,  USA}
\author{S.~Desai}
\affiliation{Department of Physics, IIT Hyderabad, Kandi, Telangana 502285, India}
\author{H.~T.~Diehl}
\affiliation{Fermi National Accelerator Laboratory, P. O. Box 500, Batavia, IL 60510, USA}
\author{J.~P.~Dietrich}
\affiliation{Excellence Cluster Universe, Boltzmannstr.\ 2, 85748 Garching, Germany}
\affiliation{Faculty of Physics, Ludwig-Maximilians-Universit\"at, Scheinerstr. 1, 81679 Munich, Germany}
\author{A.~E.~Evrard}
\affiliation{Department of Astronomy, University of Michigan, Ann Arbor, MI 48109, USA}
\affiliation{Department of Physics, University of Michigan, Ann Arbor, MI 48109, USA}
\author{B.~Flaugher}
\affiliation{Fermi National Accelerator Laboratory, P. O. Box 500, Batavia, IL 60510, USA}
\author{P.~Fosalba}
\affiliation{Institut de Ci\`encies de l'Espai, IEEC-CSIC, Campus UAB, Carrer de Can Magrans, s/n,  08193 Bellaterra, Barcelona, Spain}
\author{J.~Frieman}
\affiliation{Fermi National Accelerator Laboratory, P. O. Box 500, Batavia, IL 60510, USA}
\affiliation{Kavli Institute for Cosmological Physics, University of Chicago, Chicago, IL 60637, USA}
\author{J.~Garc\'ia-Bellido}
\affiliation{Instituto de Fisica Teorica UAM/CSIC, Universidad Autonoma de Madrid, 28049 Madrid, Spain}
\author{E.~Gaztanaga}
\affiliation{Institut de Ci\`encies de l'Espai, IEEC-CSIC, Campus UAB, Carrer de Can Magrans, s/n,  08193 Bellaterra, Barcelona, Spain}
\author{T.~Giannantonio}
\affiliation{Institute of Astronomy, University of Cambridge, Madingley Road, Cambridge CB3 0HA, UK}
\affiliation{Kavli Institute for Cosmology, University of Cambridge, Madingley Road, Cambridge CB3 0HA, UK}
\author{D.~Gruen}
\affiliation{Kavli Institute for Particle Astrophysics \& Cosmology, P. O. Box 2450, Stanford University, Stanford, CA 94305, USA}
\affiliation{SLAC National Accelerator Laboratory, Menlo Park, CA 94025, USA}
\author{R.~A.~Gruendl}
\affiliation{Department of Astronomy, University of Illinois, 1002 W. Green Street, Urbana, IL 61801, USA}
\affiliation{National Center for Supercomputing Applications, 1205 West Clark St., Urbana, IL 61801, USA}
\author{J.~Gschwend}
\affiliation{Laborat\'orio Interinstitucional de e-Astronomia - LIneA, Rua Gal. Jos\'e Cristino 77, Rio de Janeiro, RJ - 20921-400, Brazil}
\affiliation{Observat\'orio Nacional, Rua Gal. Jos\'e Cristino 77, Rio de Janeiro, RJ - 20921-400, Brazil}
\author{G.~Gutierrez}
\affiliation{Fermi National Accelerator Laboratory, P. O. Box 500, Batavia, IL 60510, USA}
\author{K.~Honscheid}
\affiliation{Center for Cosmology and Astro-Particle Physics, The Ohio State University, Columbus, OH 43210, USA}
\affiliation{Department of Physics, The Ohio State University, Columbus, OH 43210, USA}
\author{D.~J.~James}
\affiliation{Astronomy Department, University of Washington, Box 351580, Seattle, WA 98195, USA}
\affiliation{Cerro Tololo Inter-American Observatory, National Optical Astronomy Observatory, Casilla 603, La Serena, Chile}
\author{T.~Jeltema}
\affiliation{Santa Cruz Institute for Particle Physics, Santa Cruz, CA 95064, USA}
\author{K.~Kuehn}
\affiliation{Australian Astronomical Observatory, North Ryde, NSW 2113, Australia}
\author{S.~Kuhlmann}
\affiliation{Argonne National Laboratory, 9700 South Cass Avenue, Lemont, IL 60439, USA}
\author{O.~Lahav}
\affiliation{Department of Physics \& Astronomy, University College London, Gower Street, London, WC1E 6BT, UK}
\author{M.~Lima}
\affiliation{Departamento de F\'isica Matem\'atica, Instituto de F\'isica, Universidade de S\~ao Paulo, CP 66318, S\~ao Paulo, SP, 05314-970, Brazil}
\affiliation{Laborat\'orio Interinstitucional de e-Astronomia - LIneA, Rua Gal. Jos\'e Cristino 77, Rio de Janeiro, RJ - 20921-400, Brazil}
\author{M.~A.~G.~Maia}
\affiliation{Laborat\'orio Interinstitucional de e-Astronomia - LIneA, Rua Gal. Jos\'e Cristino 77, Rio de Janeiro, RJ - 20921-400, Brazil}
\affiliation{Observat\'orio Nacional, Rua Gal. Jos\'e Cristino 77, Rio de Janeiro, RJ - 20921-400, Brazil}
\author{M.~March}
\affiliation{Department of Physics and Astronomy, University of Pennsylvania, Philadelphia, PA 19104, USA}
\author{J.~L.~Marshall}
\affiliation{George P. and Cynthia Woods Mitchell Institute for Fundamental Physics and Astronomy, and Department of Physics and Astronomy, Texas A\&M University, College Station, TX 77843,  USA}
\author{P.~Martini}
\affiliation{Center for Cosmology and Astro-Particle Physics, The Ohio State University, Columbus, OH 43210, USA}
\affiliation{Department of Astronomy, The Ohio State University, Columbus, OH 43210, USA}
\author{F.~Menanteau}
\affiliation{Department of Astronomy, University of Illinois, 1002 W. Green Street, Urbana, IL 61801, USA}
\affiliation{National Center for Supercomputing Applications, 1205 West Clark St., Urbana, IL 61801, USA}
\author{R.~Miquel}
\affiliation{Instituci\'o Catalana de Recerca i Estudis Avan\c{c}ats, E-08010 Barcelona, Spain}
\affiliation{Institut de F\'{\i}sica d'Altes Energies (IFAE), The Barcelona Institute of Science and Technology, Campus UAB, 08193 Bellaterra (Barcelona) Spain}
\author{R.~C.~Nichol}
\affiliation{Institute of Cosmology \& Gravitation, University of Portsmouth, Portsmouth, PO1 3FX, UK}
\author{A.~A.~Plazas}
\affiliation{Jet Propulsion Laboratory, California Institute of Technology, 4800 Oak Grove Dr., Pasadena, CA 91109, USA}
\author{A.~K.~Romer}
\affiliation{Department of Physics and Astronomy, Pevensey Building, University of Sussex, Brighton, BN1 9QH, UK}
\author{E.~S.~Rykoff}
\affiliation{Kavli Institute for Particle Astrophysics \& Cosmology, P. O. Box 2450, Stanford University, Stanford, CA 94305, USA}
\affiliation{SLAC National Accelerator Laboratory, Menlo Park, CA 94025, USA}
\author{E.~Sanchez}
\affiliation{Centro de Investigaciones Energ\'eticas, Medioambientales y Tecnol\'ogicas (CIEMAT), Madrid, Spain}
\author{V.~Scarpine}
\affiliation{Fermi National Accelerator Laboratory, P. O. Box 500, Batavia, IL 60510, USA}
\author{R.~Schindler}
\affiliation{SLAC National Accelerator Laboratory, Menlo Park, CA 94025, USA}
\author{M.~Schubnell}
\affiliation{Department of Physics, University of Michigan, Ann Arbor, MI 48109, USA}
\author{I.~Sevilla-Noarbe}
\affiliation{Centro de Investigaciones Energ\'eticas, Medioambientales y Tecnol\'ogicas (CIEMAT), Madrid, Spain}
\author{M.~Smith}
\affiliation{School of Physics and Astronomy, University of Southampton,  Southampton, SO17 1BJ, UK}
\author{M.~Soares-Santos}
\affiliation{Fermi National Accelerator Laboratory, P. O. Box 500, Batavia, IL 60510, USA}
\author{F.~Sobreira}
\affiliation{Instituto de F\'isica Gleb Wataghin, Universidade Estadual de Campinas, 13083-859, Campinas, SP, Brazil}
\affiliation{Laborat\'orio Interinstitucional de e-Astronomia - LIneA, Rua Gal. Jos\'e Cristino 77, Rio de Janeiro, RJ - 20921-400, Brazil}
\author{E.~Suchyta}
\affiliation{Computer Science and Mathematics Division, Oak Ridge National Laboratory, Oak Ridge, TN 37831}
\author{M.~E.~C.~Swanson}
\affiliation{National Center for Supercomputing Applications, 1205 West Clark St., Urbana, IL 61801, USA}
\author{G.~Tarle}
\affiliation{Department of Physics, University of Michigan, Ann Arbor, MI 48109, USA}
\author{D.~L.~Tucker}
\affiliation{Fermi National Accelerator Laboratory, P. O. Box 500, Batavia, IL 60510, USA}
\author{V.~Vikram}
\affiliation{Argonne National Laboratory, 9700 South Cass Avenue, Lemont, IL 60439, USA}
\author{A.~R.~Walker}
\affiliation{Cerro Tololo Inter-American Observatory, National Optical Astronomy Observatory, Casilla 603, La Serena, Chile}
\author{J.~Weller}
\affiliation{Excellence Cluster Universe, Boltzmannstr.\ 2, 85748 Garching, Germany}
\affiliation{Max Planck Institute for Extraterrestrial Physics, Giessenbachstrasse, 85748 Garching, Germany}
\affiliation{Universit\"ats-Sternwarte, Fakult\"at f\"ur Physik, Ludwig-Maximilians Universit\"at M\"unchen, Scheinerstr. 1, 81679 M\"unchen, Germany}

\collaboration{DES Collaboration}
\begin{abstract} 
We present the methodology for and detail the implementation of the Dark Energy Survey (DES) 3x2pt DES Year 1 (Y1) analysis, which combines configuration-space two-point statistics from three different cosmological probes: cosmic shear, galaxy--galaxy lensing, and galaxy clustering, using data from the first year of DES observations. We have developed two independent modeling pipelines and describe the code validation process. We derive expressions for analytical real-space multi-probe covariances, and describe their validation with numerical simulations. We stress-test the inference pipelines in simulated likelihood analyses that vary 6--7 cosmology parameters plus 20 nuisance parameters and precisely resemble the analysis to be presented in the DES 3x2pt analysis paper, using a variety of simulated input data vectors with varying assumptions.

We find that any disagreement between pipelines leads to changes in assigned likelihood $\Delta \chi^2 \le 0.045$ with respect to the statistical error of the DES Y1 data vector. We also find that angular binning and survey mask do not impact our analytic covariance at a significant level. We determine lower bounds on scales used for analysis of galaxy clustering (8 Mpc$~h^{-1}$) and galaxy--galaxy lensing (12 Mpc$~h^{-1}$) such that the impact of modeling uncertainties in the non-linear regime is well below statistical errors, and show that our analysis choices are robust against a variety of systematics. 
These tests demonstrate that we have a robust analysis pipeline that yields unbiased cosmological parameter inferences for the flagship 3x2pt DES Y1 analysis. We emphasize that the level of independent code development and subsequent code comparison as demonstrated in this paper is necessary to produce credible constraints from increasingly complex multi-probe analyses of current data. 
\end{abstract}
\maketitle

\section{Introduction}
\label{sec:intro}

Ongoing photometric surveys, such as Kilo-Degree Survey (KiDS\footnote{http://www.astro-wise.org/projects/KIDS/}), Hyper Suprime Cam (HSC\footnote{http://www.naoj.org/Projects/HSC/HSCProject.html}), and the Dark Energy Survey (DES\footnote{www.darkenergysurvey.org/}) enable detailed measurements of the late-time Universe and powerful tests of the nature of cosmic acceleration and General Relativity.
Even more powerful measurements will be made in the early 2020s by even larger experiments, e.g., the Large Synoptic Survey Telescope (LSST\footnote{http://www.lsst.org/lsst}), Euclid\footnote{sci.esa.int/euclid/} and the Wide-Field Infrared Survey Telescope (WFIRST\footnote{http://wfirst.gsfc.nasa.gov/}). 
  
The complete DES will map $\sim$ 5000 deg$^2$ \citep{des} and constrain cosmology with multiple probes, including cosmic shear, galaxy--galaxy lensing, galaxy clustering, Baryon Acoustic Oscillations (BAO), galaxy cluster number counts, and Type Ia supernovae (SNIa). Such a multi-probe approach is the most promising route to uncovering and characterizing any new cosmological physics. Tension between the individual probes can lead to insight into new physical concepts or hint at neglected systematic effects. If the individual probes are consistent with each other, their joint analysis will lead to a substantial gain in information in the joint parameter space of cosmology and systematic effects.  
\begin{figure*}
  \includegraphics[width=0.45\textwidth]{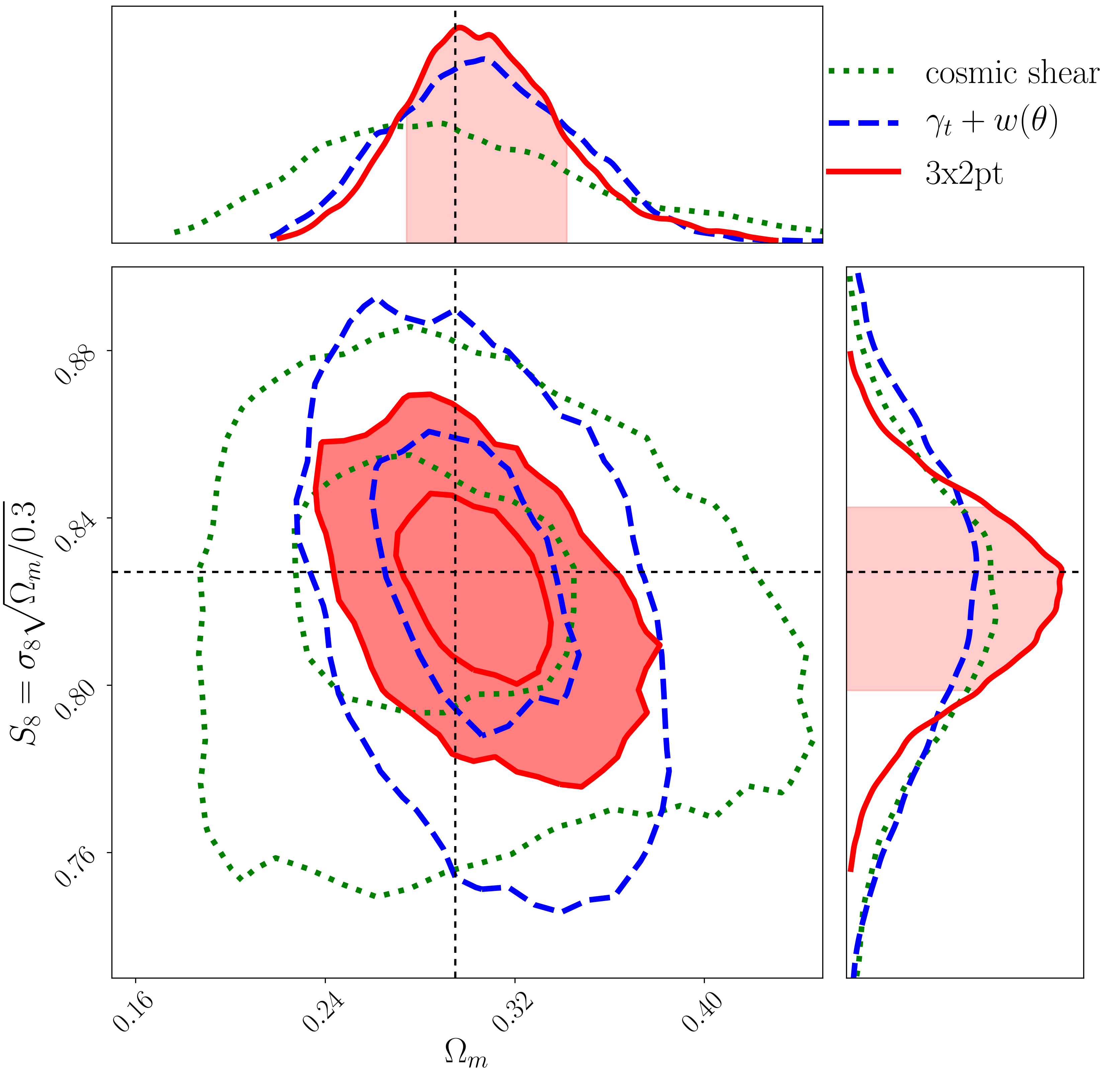}
  \includegraphics[width=0.45\textwidth]{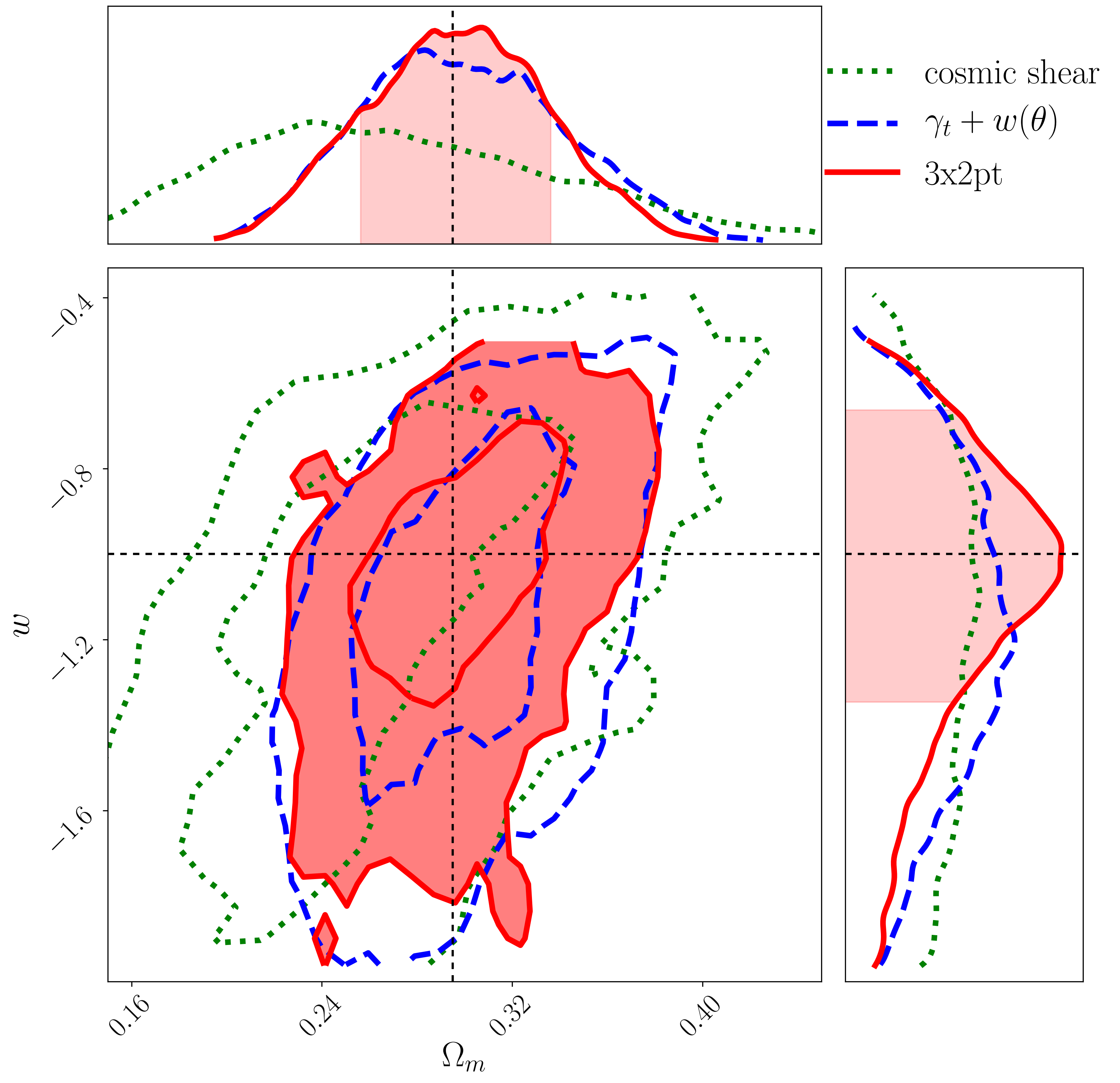}
\caption{\emph{Left}: 1$\sigma$ and 2$\sigma$ contours show the forecast marginalized constraints on the matter density $\Omega_\mathrm{m}$ and amplitude parameter $S_8 = \sigma_8\sqrt{\Omega_\mathrm{m}/0.3}$ from simulated analyses of the DES-Y1 cosmic shear (green, dotted), galaxy clustering + galaxy--galaxy lensing (blue, dashed), and 3x2pt data vectors assuming the baseline analysis choices developed in this paper for flat $\Lambda$CDM cosmology. The black dashed lines indicate the fiducial cosmology. \emph{Right}: Contours are as in the left panel, here showing $\Omega_\mathrm{m}$ vs. the dark energy equation of state parameter $w$, assuming a flat $w$CDM cosmology.}
 \label{fig:singlevsmulti}
\end{figure*}

A number of studies have combined individual large-scale structure probes with SNIa or Cosmic Microwave Background (CMB) measurements \citep[e.g.,][]{Betoule2014, DES15WL, planck15, alam16}. In both of these cases, the information from the two sets of probes is largely uncorrelated. However, the other major cosmological probes from large galaxy surveys are highly correlated with each other in that they are tracers of the same underlying density field, and in that they share common systematic effects. Consequently, a multi-probe analysis based on correlated photometric probes can no longer simply combine the optimal versions of individual analyses. Instead, to take full advantage of the power of combining probes of large-scale structure, one must build a tailored analysis pipeline that can model cosmological observables and their correlated systematics consistently. In addition to this modeling framework, multi-probe analyses require the ability to compute joint covariance matrices that properly account for the cross-correlation of various observables. In this work, we present the development and validation of this framework and these covariances for the combined probes analysis of DES Y1 data.

To date, these complications have prohibited or at least severely limited multi-probe analyses that combine different tracers of the Universe's large-scale structure (LSS) from photometric data sets. The potential gain in constraining power from successfully implementing a photometric multi-probe analysis has however been forecast with different levels of complexity. For example, \citet{ber09} gave a detailed description of a Fisher matrix analysis of galaxy clustering, cosmic shear, and galaxy--galaxy lensing. Similar analyses were presented in \cite{job10} and \cite{yos12}, where the latter considered number counts of galaxy clusters instead of cosmic shear. All three analyses used Gaussian covariances, where Gaussian means that connected higher-order moments of the density field are not included in the covariance computation. However, covariance terms that arise from these higher-order moments can significantly impact error bars \citep{coh01,taj09,sht09}; these terms were included in the analyses of \cite{eks14, tas14, pkd15,Lacasa:2016pxq} all of which vary in terms of the probes considered. 

\citet{cosmolike} simulated joint analyses of cosmic shear, galaxy--galaxy lensing, galaxy clustering, photometric BAO, galaxy cluster number counts, and galaxy cluster weak lensing. This analysis included all cross-correlations among probes, derived an analytical expression for non-Gaussian covariances, and simultaneously modeled uncertainties from photo-$z$ and galaxy shape measurements, galaxy bias models, cluster-mass observable relation, and galaxy intrinsic alignments.  
While the aforementioned forecasts demonstrate the potential of a multi-probe analysis with photometric data, few such analyses have been performed \citep[see e.g.,][]{ltb12,cvm13,msb13,kwan16,nra17,KiDS32pt}. \citet{msb13} constrain cosmological parameters from large-scale galaxy--galaxy lensing and galaxy clustering in the Sloan Digital Sky Survey. The analyses \citep{kwan16,nra17} use galaxy--galaxy lensing and galaxy clustering from DES Science Verification data and account for cross-correlations through Fourier-space Gaussian or Jackknife covariances. \citet{KiDS32pt} present a joint analysis of cosmic shear, galaxy--galaxy lensing and galaxy clustering using power spectrum measurements, combining weak lensing from $\sim$450 deg$^2$ of KiDS \citep{KiDS450} with a spectroscopic galaxy sample from the Galaxies And Mass Assembly (GAMA) survey in $\sim$180 deg$^2$ of KiDS--GAMA overlap area.

The DES-Year1 3x2pt key project \citep[][Y1KP hereafter]{keypaper} takes photometric multi-probe analyses to the next level: In this paper we demonstrate the ability of DES to conduct a joint cosmic shear, galaxy--galaxy lensing, and galaxy clustering analysis. We verify that the precision of our inference pipelines is sufficient for the statistical constraining power of the 1321 deg$^2$ Y1 area footprint. Furthermore, we determine scale cuts to ensure that the impact of modeling uncertainties in the non-linear regime is well below statistical errors of the analysis. Throughout this paper, we focus on constraints on the matter density parameter $\Omega_\mathrm{m}$, the parameter $S_8=\sigma_8\sqrt{\Omega_\mathrm{m}/0.3}$ which measures the amplitude of structure growth, and the dark energy equation of state parameter $w$ (assumed constant in time), as these are the main results of the Y1KP analysis. Interesting constraints on the time evolution of the dark energy equation of state $w_a$ will be the goal of future analyses that use the full DES survey area.

The gain in information when combining the three two-point functions is illustrated in Fig. \ref{fig:singlevsmulti}, which compares the 3x2pt analysis with a cosmic shear only and a galaxy--galaxy lensing plus galaxy clustering analysis using the same data. These simulated results correspond to the baseline Y1KP likelihood analysis including all systematics and scale cuts; the only difference is that the data vector is not computed from the measurements in the DES catalogs but is generated from our modeling frameworks using a fiducial set of parameters (see Table \ref{tab:params}). These priors on observational systematic effects reflect the current state of the Y1KP analyses. The final cosmology analysis may use slightly different priors, which will not alter the conclusions of this paper.

It is the main goal of the present work to motivate and validate the likelihood analysis details of the Y1KP measurement \cite{keypaper}. We describe the methodology, including details of the cosmological modeling, covariance derivation, and systematics mitigation through scale-cuts and marginalization. 
We have developed two independent implementations for the cosmological likelihoods, building on the \textsc{CosmoSIS} \citep{zpj15} and the \textsc{CosmoLike} \citep{cosmolike} modeling frameworks.  We conduct a detailed code comparison between these two independent cosmological likelihood implementations and demonstrate that they agree extremely well. This comparison was a long-term, core project of the present work and the importance of such a parallel code implementation and subsequent comparison to ensure the accuracy of our analysis is hard to overstate.

The connection of this paper to the Y1KP data papers can be illustrated by Eq.~(\ref{eq:like}), which is at the core of the Y1KP parameter inference, i.e. computing the likelihood of the data $\D$ given a point in cosmological and systematics parameter space $\p$.   
\be
\label{eq:like}
\like (\D| \p) \propto \exp \biggl( -\frac{1}{2} \left[ \left(\D -\M(\p)\right)^t \, \matC^{-1} \, \left(\D-\M(\p)\right) \right]  \biggr) \,.
\ee
The data vector $\D$ is delivered to the Y1KP through multiple essential DES papers \citep{y1gold,xcorrtechnique,xcorr,redmagicpz,photoz, wthetapaper,gglpaper,shearcorr,shearcat}, which detail the value-added galaxy catalog, redshift distributions, survey mask, systematics priors, galaxy number densities for the lens and source samples, and measurements of the two-point correlation functions. Based on this information it is the task of this paper to implement the modeling framework that will allow the Y1KP to obtain the model vector $\M$, and to provide the capability to robustly compute a covariance matrix $\matC$. The tests of the modeling and inference accuracy presented in this paper throughout are based on simulated analyses of synthetic data vectors, for which the input cosmology and systematic contaminations are known exactly. \citet{simspaper} validate this modeling and inference framework on mock catalogs generated to mimic many aspects of the Y1 data sets \citep{buzzard}.

This paper is structured as follows. In Sec.~\ref{sec:model} we describe the equations implemented in our analysis pipelines, and the code comparison. Section~\ref{sec:cov} details the covariance modeling and validation effort. Section~\ref{sec:choices} stress-tests our pipelines through a variety of simulated analyses, which determines our analysis choices and demonstrates robustness against systematic effects. In Sec. \ref{sec:samplers} we describe optimal settings for our likelihood samplers. In Sec. \ref{sec:neutrinos} we test an extension of our analysis frameworks, namely the inclusion of massive neutrinos. We conclude in Sect. \ref{sec:conc}.  Further details of the code comparison are given in Appendix \ref{sec:codecomp}.

\section{Models for Multi-Probe Summary Statistics}
The DES 3x2pt data vector consists of angular galaxy clustering, galaxy--galaxy lensing, and cosmic shear two-point function measurements. This section describes the theoretical baseline model for the data vector, and the validation of our numerical implementation.  
\label{sec:model}
\subsection{Angular two-point functions}
\label{sec:cosmo}
The DES-Y1 3x2pt correlation functions are measured as the auto- and cross-correlations of two galaxy catalogs: The first catalog contains the positions of ``lens" galaxies selected using the redMaGiC algorithm \citep{RRA15}, which are used for clustering measurements and as lens galaxies for galaxy--galaxy lensing. The second catalog contains the positions and shape estimates from the weak lensing ``source" galaxy sample, which are used for cosmic shear measurements and as source galaxies for galaxy--galaxy lensing. The redMaGiC sample selection and redshift calibration are described in \citet{wthetapaper,redmagicpz}; the selection of the weak lensing source sample from the DES-Y1 gold catalog \citep{y1gold} and the shear catalog are  described in \citet{shearcat}, and the source redshift estimates are described \citet{photoz}, respectively. We summarize here the specifications of the Y1KP data, which we use as input for the simulated likelihood analyses presented in this paper.

\begin{figure}
\includegraphics[width=8.5cm]{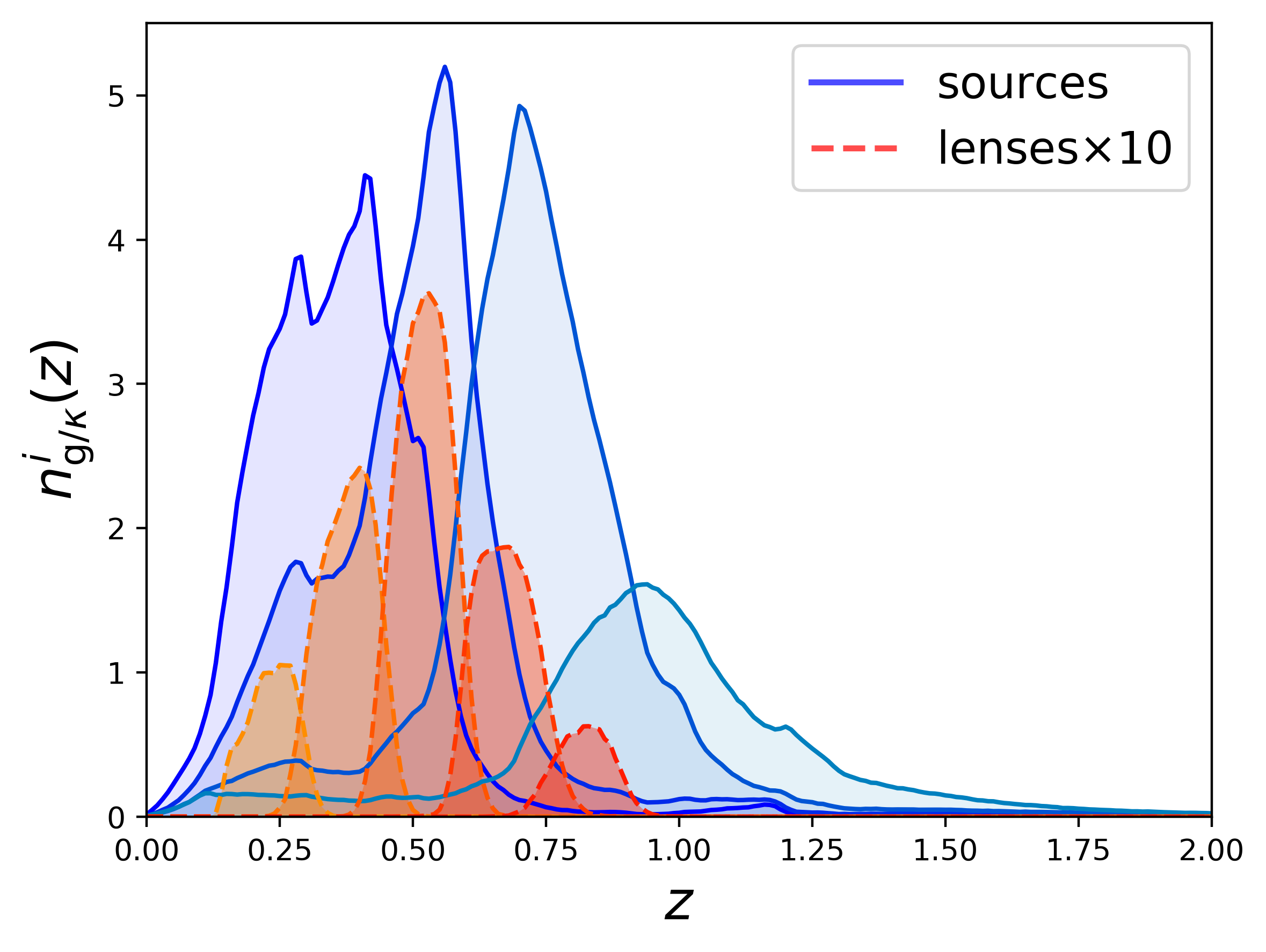}
\caption{Estimated redshift distributions of the redMaGiC lens galaxy sample (dashed lines) and the \textsc{metacal} source galaxy sample (solid lines) for the Y1KP analysis. The  lens and source galaxies are split into five and four tomography bins respectively. See \citep{xcorr,xcorrtechnique,redmagicpz,photoz} for details.}
\label{fig:nz}
\end{figure}
\renewcommand{\arraystretch}{1.3}
\begin{table}
\caption{Parameters of the baseline model: fiducial values, flat priors (min, max), and Gaussian priors ($\mu$, $\sigma$). See Sect.~\ref{sec:sys} for a description of the systematics parameters, and Y1KP for a discussion of the cosmology parameters and priors. The priors on observational systematic effects reflect the current state of the Y1KP analyses to be detailed in \citet[][redMaGiC photo-z]{redmagicpz}, \citet[][source photo-z]{photoz}, \citet[][shear calibration]{shearcat}. The final analyses may use slightly different priors, which will not alter the conclusions of this paper.}
\begin{center}
\begin{tabular*}{0.48\textwidth}{@{\extracolsep{\fill}}| c c c |}
\hline
\hline
Parameter & Fiducial & Prior \\  
\hline 
\hline 
\multicolumn{3}{|c|}{\tbf{Cosmology} \citep{keypaper}} \\
$\om$ & 0.295 &  flat (0.1, 0.9)  \\ 
$A_\mathrm{s}/10^{-9}$ &2.26 &  flat (0.5, 5.0)  \\ 
$\ns$ & 0.968 & flat (0.87, 1.07)  \\
$w$ &  -1.0 &   flat (-2.0, -1/3)   \\
$\omb$ &  0.044 &  flat (0.03, 0.07)  \\
$h_0$  & 0.6881 &  flat (0.55, 0.91)   \\
$\Omega_\nu h^2$ & $6.16\times 10^{-4}$ & fixed; varied in Sect.~\ref{sec:neutrinos}\\
$\Omega_\mathrm{K}$ & $0$ & fixed \\
\hline
\multicolumn{3}{|c|}{\tbf{Galaxy Bias}} \\
$b_1^1$ & 1.45  & flat (0.8, 3.0) \\
$b_1^2$ & 1.55  &flat (0.8, 3.0) \\
$b_1^3$ & 1.65 & flat (0.8, 3.0) \\
$b_1^4$ & 1.8 & flat (0.8, 3.0) \\
$b_1^5$ & 2.0 & flat (0.8, 3.0) \\
\hline
\multicolumn{3}{|c|}{\tbf{redMaGiC Photo-$z$ }} \\ 
$\Delta_\mr{z,g}^1 $ & 0.002 & Gauss (0.002, 0.007) \\
$\Delta_\mr{z,g}^2 $ & 0.001 & Gauss (0.001, 0.007) \\
$\Delta_\mr{z,g}^3 $ & 0.003 & Gauss (0.003, 0.007) \\
$\Delta_\mr{z,g}^4 $ & 0.0 & Gauss (0.0, 0.01) \\
$\Delta_\mr{z,g}^5 $ & 0.0 & Gauss (0.0, 0.01) \\
\hline
\multicolumn{3}{|c|}{\textsc{metacal} \tbf{Source Photo-$z$ }} \\ $\Delta_\mr{z,\kappa}^1 $ & 0.000 & Gauss (0.000, 0.018) \\
$\Delta_\mr{z,\kappa}^2 $ & -0.014 & Gauss (-0.014, 0.013) \\
$\Delta_\mr{z,\kappa}^3 $ & 0.014 & Gauss (0.014, 0.011) \\
$\Delta_\mr{z,\kappa}^4 $ & 0.033 & Gauss (0.033, 0.022) \\
\hline
\multicolumn{3}{|c|}{\textsc{metacal} \tbf{Shear Calibration}} \\
$m^i $ & 0.013 & Gauss (0.013, 0.021)\\
\hline
\multicolumn{3}{|c|}{\tbf{Intrinsic Alignments}} \\
$A_{\mathrm{IA,0}} $ & 0.0 & flat (-5.0, 5.0)\\
$\alpha_{\mathrm{IA}} $ & 0.0 & flat (-5.0, 5.0)\\
$z_0$& 0.62 & fixed\\
\hline
\end{tabular*}
\end{center}
\label{tab:params}
\end{table}
\renewcommand{\arraystretch}{1.0}

\paragraph*{Source galaxies} We use the redshift distribution of the \textsc{metacal} \citep[see][for details of the algorithm]{metacal1,metacal2} shear catalog described in \cite{shearcat}.  This includes 5.2 galaxies/arcmin$^2$, split into 4 tomography bins. These are shown as solid lines in Fig.~\ref{fig:nz}, with effective number densities of 1.5, 1.5, 1.6, 0.8 galaxies/arcmin$^2$, for the 4 bins respectively.
\paragraph*{Lens galaxies } The redMaGiC lens galaxy sample is described in \cite{wthetapaper} and split into 5 tomographic bins, which are shown as dashed lines in Fig.~\ref{fig:nz}, with number densities of 0.013, 0.03, 0.05, 0.03, 0.009 galaxies/arcmin$^2$, for the 5 bins respectively.
\paragraph*{Galaxy--galaxy lensing} We consider all combinations of lens and source bins for the galaxy--galaxy lensing correlation functions. While galaxy--galaxy lensing requires the source galaxies to be located at higher redshift than the lens galaxies, the signals from all tomography bin combinations contribute to the self-calibration of photometric redshifts, intrinsic alignments, and other systematic effects.\newline

We denote the projected (angular) density contrast of redMaGiC galaxies in redshift bin $i$ by $\delta_{\mathrm{g}}^i$, the convergence field of source tomography bin $j$ as $\kappa^j$, the redshift distribution of the redMaGiC/source galaxy sample in tomography bin $i$ as $n_{\mathrm{g}/\kappa}^i(z)$, and the angular number densities of galaxies in this redshift bin as 
\be
\bar{n}_{\mathrm{g}/\kappa}^i = \int dz\; n_{\mathrm{g}/\kappa}^i(z)\,.
\ee
The radial weight function for clustering in terms of the comoving radial distance $\chi$ is
\be
q_{\delta_{\mathrm{g}}}^i(k,\chi) = b^i\left(k,z(\chi)\right)\frac{n_{\mathrm{g}}^i(z(\chi)) }{\bar{n}_{\mathrm{g}}^i}\frac{dz}{d\chi}\,,
\ee
with $b^i(k,z(\chi))$ the galaxy bias of the redMaGiC galaxies in tomography bin $i$,
and the lensing efficiency 
\be
q_\kappa^{i}(\chi) = \frac{3 H_0^2 \Omega_m }{2 \mathrm{c}^2}\frac{\chi}{a(\chi)}\int_\chi^{\chi_{\mr h}} \mr d \chi' \frac{n_{\kappa}^{i} (z(\chi')) dz/d\chi'}{\bar{n}_{\kappa}^{i}} \frac{\chi'-\chi}{\chi'} \,,
\ee
with $H_0$ the Hubble constant, $c$ the speed of light, and $a$ the scale factor.
Under the Limber approximation, the angular power spectra for cosmic shear, galaxy-galaxy lensing, and galaxy clustering can be written as
\begin{align}
\label{eq:Cell}
\nonumber C_{\kappa \kappa}^{ij} (l) &=\!\! \int\!\! d\chi \frac{q_\kappa^i(\chi) q_\kappa^j(\chi) }{\chi^2} P_{\mathrm{NL}}\!\!\left(\frac{l+1/2}{\chi},z(\chi)\right)\\
\nonumber C_{\delta_{\mathrm{g}}\kappa}^{ij}(l) &=\!\! \int\!\! d\chi\! \frac{q_{\delta_{\mathrm{g}}}^i\!\!\left(\frac{l+1/2}{\chi},\chi\right) q_\kappa^j(\chi)}{\chi^2} P_{\mathrm{NL}}\!\left(\frac{l+1/2}{\chi},z(\chi)\right)\\
C_{\delta_{\mathrm{g}}\delta_{\mathrm{g}}}^{ij}(l) &=\!\! \int\!\! d\chi\! \frac{q_{\delta_{\mathrm{g}}}^i\!\!\left(\!\frac{l+1/2}{\chi},\chi\right)q_{\delta_{\mathrm{g}}}^j\!\left(\frac{l+1/2}{\chi},\chi\right)}{\chi^2} P_{\mathrm{NL}}\!\!\left(\frac{l+1/2}{\chi},z(\chi)\!\!\right)\,
\end{align}
with $P_{\mathrm{NL}}(k,z)$ the non-linear matter power spectrum at wave vector $k$ and redshift $z$.

The angular two-point clustering correlation function $w$ is computed from the angular power spectrum as
\be
w^{i}(\theta) = \sum_l \frac{2l+1}{4\pi} P_l\left(\cos(\theta)\right)\,C_{\delta_{\mathrm{g}}\delta_{\mathrm{g}}}^{ii}(l)\,,
\ee
with $P_l(x)$ the Legendre polynomial of order $l$. We restricted $w$ to auto-correlations within each tomography bin, as the cross-correlations are used in the redshift validation of the redMaGiC sample \citep{redmagicpz} and are not included in the data vector for the cosmology analysis.

We compute the galaxy--galaxy lensing two-point function $\gamma_{\mathrm t}$ and the cosmic shear two-point functions $\xi_\pm$ using the flat-sky approximation
\begin{align}
\gamma_{\mathrm{t}}^{ij}(\theta) &=\int \frac{dl\, l}{2\pi} J_2(l\theta) C_{\delta_{\mathrm{g}}\kappa}^{ij}(l) \,,\\
\xi_{+/-}^{ij}(\theta) &= \int \frac{dl\, l}{2\pi} J_{0/4}(l\theta) C_{\kappa \kappa}^{ij}(l)\,,
\label{eq:xi}
\end{align}
with $J_n(x)$ the $n$-th order Bessel function of the first kind. We verified that differences between the flat-sky approximation and full-sky calculation for $\gamma_{\mathrm t}$ and $\xi_\pm$ \citep{KKS97} are negligible compared to the DES-Y1 statistical uncertainties, in agreement with \cite{Kilbinger17}.

All correlation functions are measured in 20 logarithmically spaced angular bins over the range $2 \farcm 5 < \theta < 250'$. We evaluate correlation functions in angular bin $[\theta_{\mathrm{min,i}},\theta_{\mathrm{max,i}}]$ at the area-weighted bin center $\bar{\theta}_i$,
\be
\bar{\theta}_i =  \frac{\int_{\theta_{\mathrm{min,i}}}^{\theta_{\mathrm{max,i}}} d\theta\, 2\pi \theta\,\theta} {\int_{\theta_{\mathrm{min,i}}}^{\theta_{\mathrm{max,i}}} d\theta\, 2\pi \theta} = \frac{2}{3}\frac{\left(\theta_{\mathrm{max,i}}^3-{\theta_{\mathrm{min,i}}^3}\right)}{(\theta_{\mathrm{max,i}}^2-{\theta_{\mathrm{min,i}}^2})}\,,
\ee
which is computationally faster than averaging the predicted correlation function over each bin. We verified that this approximation is sufficiently accurate given the DES-Y1 statistical uncertainties.

\subsection{Systematics} 
\label{sec:sys}
We parameterize uncertainties arising from systematics through nuisance parameters, which are summarized with their fiducial values and priors in Table \ref{tab:params}. Our baseline likelihood analyses includes the systematics models described below, and we test whether these parameterizations are sufficiently flexible for the DES-Y1 analysis in Sect.~\ref{sec:sys2}.

\paragraph*{Photometric redshift uncertainties} 
As described in \citet{photoz} and \citet{redmagicpz} the uncertainty in the redshift distribution $n$ is modeled through shift parameters $\Delta_z$,
\begin{align}
n^i_{x}(z) = \hat{n}^i_{x}\left(z-\Delta^i_{z,x}\right)\,,\;\;\; x\in\left\{\mathrm{g},\kappa\right\}\,,
\end{align}
where $\hat{n}$ denotes the estimated redshift distribution. \citet{shearcorr} and Y1KP test that this parameterization is sufficient for DES-Y1 cosmology analyses. We marginalize over one parameter for each source and lens redshift bin (nine parameters in total), using the the priors derived in \citet{photoz,redmagicpz}.
\paragraph*{Galaxy bias} The baseline model assumes an effective linear galaxy bias ($b_1$) using one parameter per lens galaxy redshift bin 
\be
b^i(k,z) = b_1^i\,,
\label{eq:bias1}
\ee
i.e. five parameters, which are marginalized over conservative flat priors.

\paragraph*{Multiplicative shear calibration} is modeled using one parameter $m^i$ per redshift bin, which affects cosmic shear and galaxy--galaxy lensing correlation functions via
\bea
\nonumber \xi_\pm^{ij}(\theta) \quad &\longrightarrow& \quad (1+m^i) \, (1+m^j) \, \xi_\pm^{ij}(\theta), \\
\gamma_t^{ij}(\theta) \quad &\longrightarrow& \quad (1+m^j) \, \gamma_t^{ij}(\theta),
\label{eq:m}
\eea
We marginalize over all four $m^i$ independently with Gaussian priors.

\paragraph*{Intrinsic galaxy alignments} (IA) are modeled using a power spectrum shape and amplitude $A(z)$. The baseline model assumes the non-linear linear alignment (NLA) model \citep{his04,brk07} for the IA power spectrum. The impact of this specific IA power spectrum model can be written as  
\bea
\label{eq:CIA}
q_\kappa^{i}(\chi) &\longrightarrow&q_\kappa^{i}(\chi) - A\left(z\left(\chi\right)\right)\frac{n^i_{\kappa}(z(\chi))}{\bar{n}^i_{\kappa}} \frac{dz}{d \chi}\,.
\eea
We model the IA amplitude assuming a power-law scaling in $(1+z)$ 
\be
A(z) = A_{\mathrm{IA,0}}\left(\frac{1+z}{1+z_0}\right)^{\alpha_{\mathrm{IA}}} \frac{C_1\rho_{\mathrm m,0}}{D(z)}\,,
\ee
with pivot redshift $z_0 =0.62$, $C_1\rho_{\mathrm{crit}} = 0.0134$ a normalization derived from SuperCOSMOS observations \citep{brk07}, and the linear growth factor $D(z)$, and marginalize over the normalization $A_{\mathrm{IA,0}}$ and power law slope $\alpha_{\mathrm{IA}}$. 

\begin{figure*}
  \includegraphics[width=0.9\textwidth]{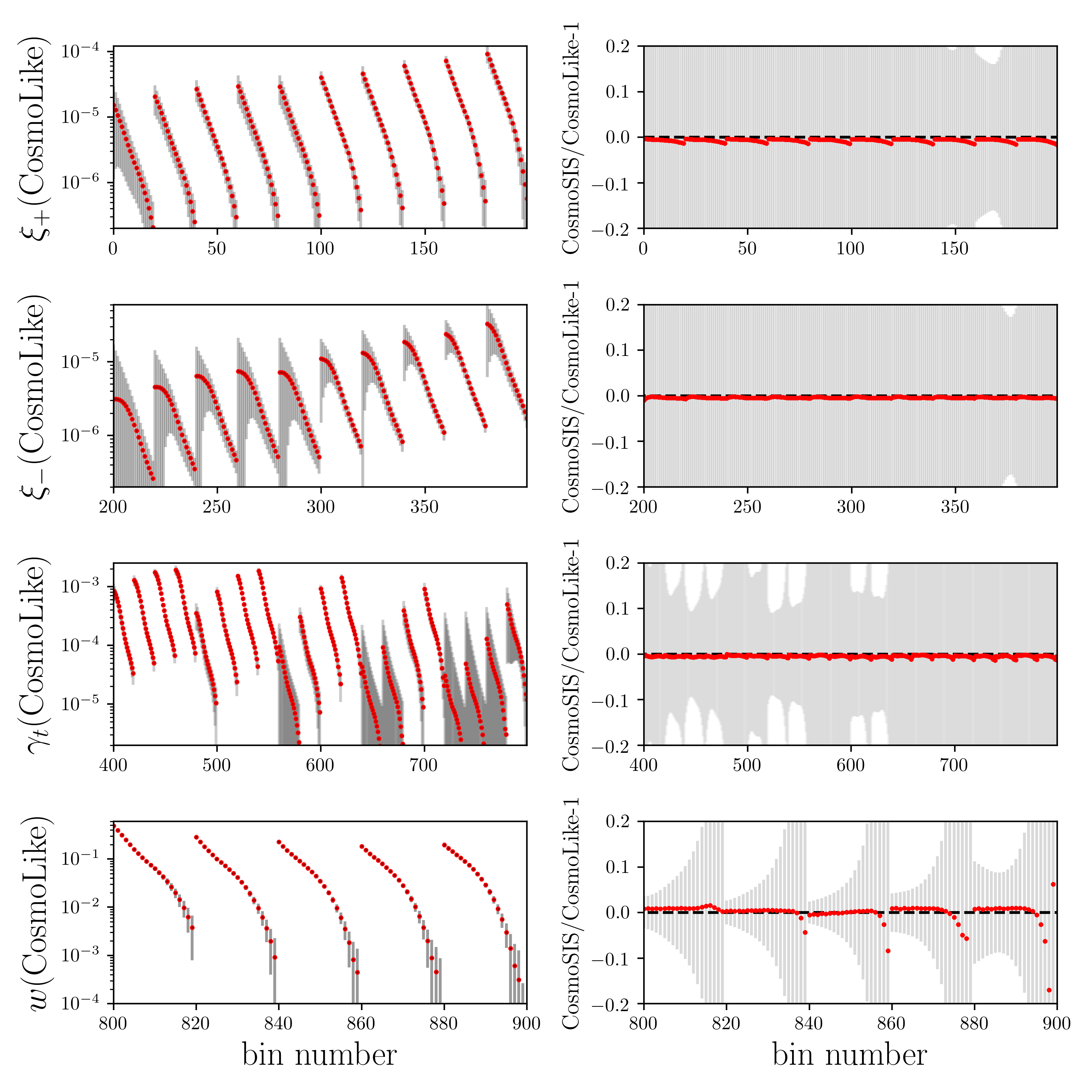}
  \caption{\emph{Left}: Model data vectors (evaluated at the fiducial parameter point), with grey error bars indicating the statistical uncertainties of the DES-Y1 analysis. From top to bottom, the panels show the $\xi_+$, $\xi_-$, $\gamma_\mathrm{t}$, and $w$ correlation functions; within each panel, angular and tomography bins are arranged along the $x$-axis, with each group of 20 data points corresponding to the 20 angular bins for each tomography bin. \emph{Right}: Fractional deviation of the data vector prediction from the two independent implementations (symbols; see text for details), using the same ordering of data points as the left panel. The grey bands show the statistical uncertainty of the DES-Y1 analysis. Within angular scale cuts of the DES-Y1 3x2pt analysis, the residual difference between these two implementations of the data vector corresponds to $\Delta \chi^2 = 0.045$.}
 \label{fig:cc1}
\end{figure*}
\begin{figure}
  \includegraphics[width=0.45\textwidth]{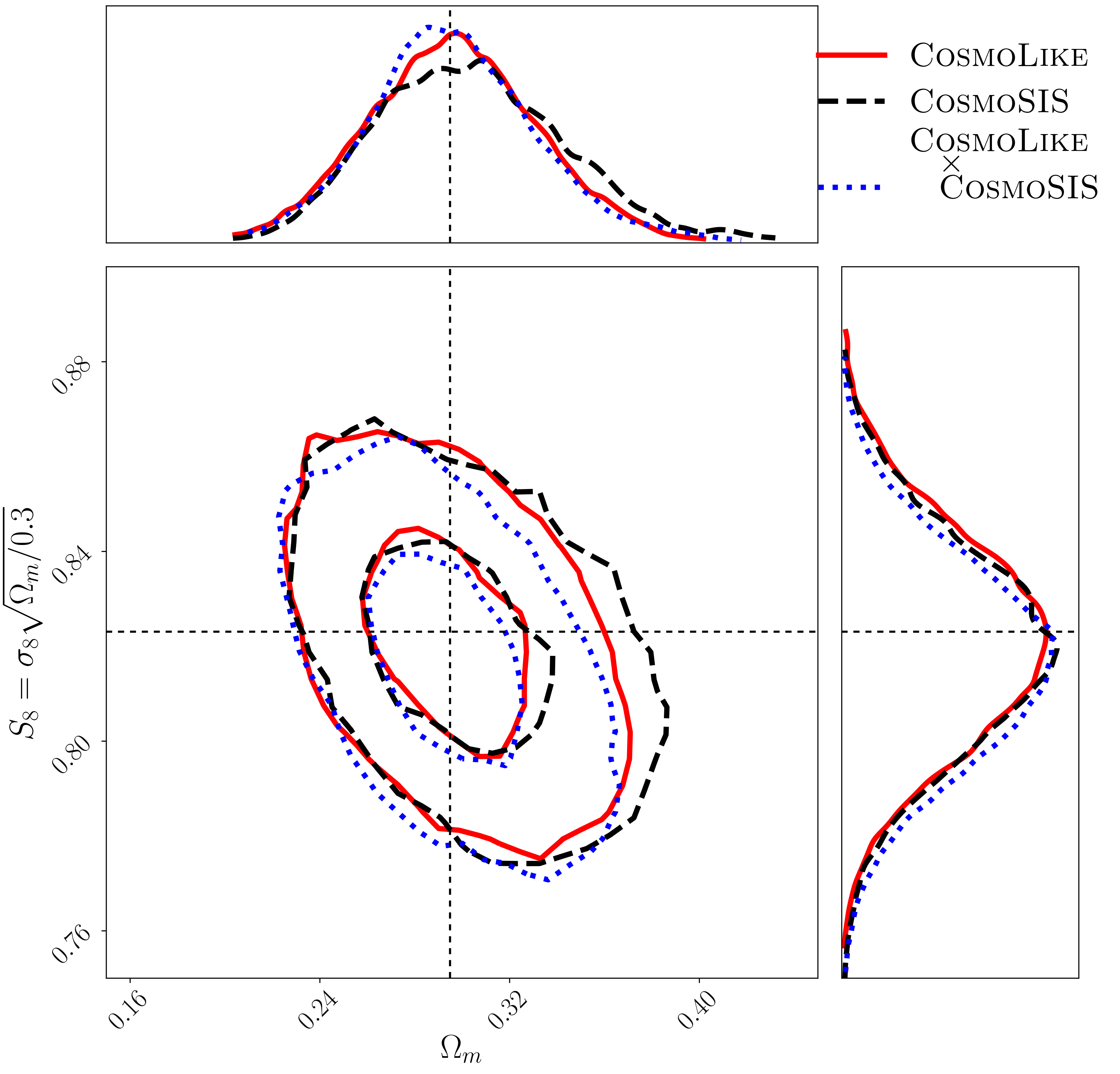}
  \caption{Comparion between parameter constraints obtained using two independent cosmological inference pipelines. The solid red/dashed black lines shows 1$\sigma$ and 2$\sigma$ parameter contours obtained using \textsc{CosmoLike}/\textsc{CosmoSIS} as modeling pipeline, respectively, with input data vectors generator by the same code. The dotted blue lines show the parameter contours obtained using \textsc{CosmoLike} as modeling pipeline, with an input data generated by \textsc{CosmoSIS}.}
 \label{fig:cc2}
\end{figure}
\subsection{Implementation}
\label{sec:code}
The baseline model for this paper assumes flat $\Lambda$CDM or $w$CDM cosmologies with the sum of neutrino masses fixed at the minimum mass consistent with bounds from oscillation measurements (see Table~\ref{tab:params}). Neutrino mass is implemented as one massive neutrino species, with the number of ultra-relativistic species fixed to get the standard model $N_\mathrm{eff} = 3.046$ at neutrino decoupling.

Among the $\Lambda$CDM/$w$CDM cosmology parameters, the DES Y1KP best constrains $\Omega_\mathrm{m}$, $S8 = \sigma_8\sqrt{\Omega_{\mathrm m}/0.3}$, and $w$. Hence we present the primary validation of the modeling pipeline and modeling parameterizations in terms of these parameters.

The correlation function model described in Sects.~\ref{sec:cosmo}-\ref{sec:sys} is implemented in two independent pipelines, the \textsc{CosmoSIS} framework and \textsc{CosmoLike}. \textsc{CosmoSIS} obtains matter power spectra through calls to the Boltzmann code \textsc{CAMB} \citep{CAMB1,CAMB2}, while \textsc{CosmoLike} calls the \textsc{CLASS} code \citep{CLASS}. Both \textsc{CAMB} and \textsc{CLASS} use the \citet{tsn12} calibration of the \textsc{halofit} fitting function for the non-linear matter power spectrum \citep{smp03}, and the \citet{Bird12} extension to include the effect of massive neutrinos on the non-linear matter power spectrum. \textsc{CosmoSIS} also calls the \textsc{nicaea} code \citep{nicaea} to compute Eqs.~(\ref{eq:xi}) via Hankel transforms.

After extensive validation, these two independent implementations are in excellent agreement over the expected Y1KP parameter space. Figure~\ref{fig:cc1} shows the fractional difference between the data vectors calculated by both pipelines at the fiducial parameters; within the angular scale cuts (c.f. Sect.~\ref{sec:choices}) the residual difference between these two implementations of the data vector corresponds to $\Delta \chi^2 = 0.045$ (using the data covariance described in Sect.~\ref{sec:cov}). In order to verify that the two implementations agree not only at the fiducial parameter point, but also in the response to parameter changes, Fig.~\ref{fig:cc2} shows the posterior likelihoods calculated from both pipelines using the model data vectors at the fiducial parameter point (from Fig.~\ref{fig:cc1}) as input. 

In order to achieve the level of agreement reported here, we compared the intermediate outputs (distances, growth factors, power spectra, etc.), as well as the posterior likelihood varying each of the 27 cosmology and nuisance parameters, holding all other 26 parameters fixed (c.f. Fig~\ref{fig:coderesponse}). The first few iterations of this comparison uncovered actual coding errors; to reach the final level of agreement, further iterations required validation of numerical implementation details, such as integration accuracy, interpolation of look-up tables, and accuracy of Hankel transform implementations in Eq.~(\ref{eq:xi}).

This comparison demonstrates that the two implementations agree sufficiently well to be run interchangeably, and that the DES-Y1 parameter constraints are robust to inaccuracies from different numerical approximation schemes.

\section{Covariance Computation}
\label{sec:cov}
As Eq.~\ref{eq:like} indicates, The covariance matrix, or more precisely its inverse, the precision matrix, is the decisive quantity that determines the errors on cosmological parameters. Obtaining precision matrices is an area of active research; methods can be broadly separated into 3 categories: estimation from numerical simulations, estimation from data directly, and analytical modeling/computation. We briefly summarize the current state of affairs as it is most relevant to our paper, however we note that our summary is far from complete.

\begin{figure*}
  \begin{minipage}[c]{0.83\textwidth}
\includegraphics[width=\textwidth]{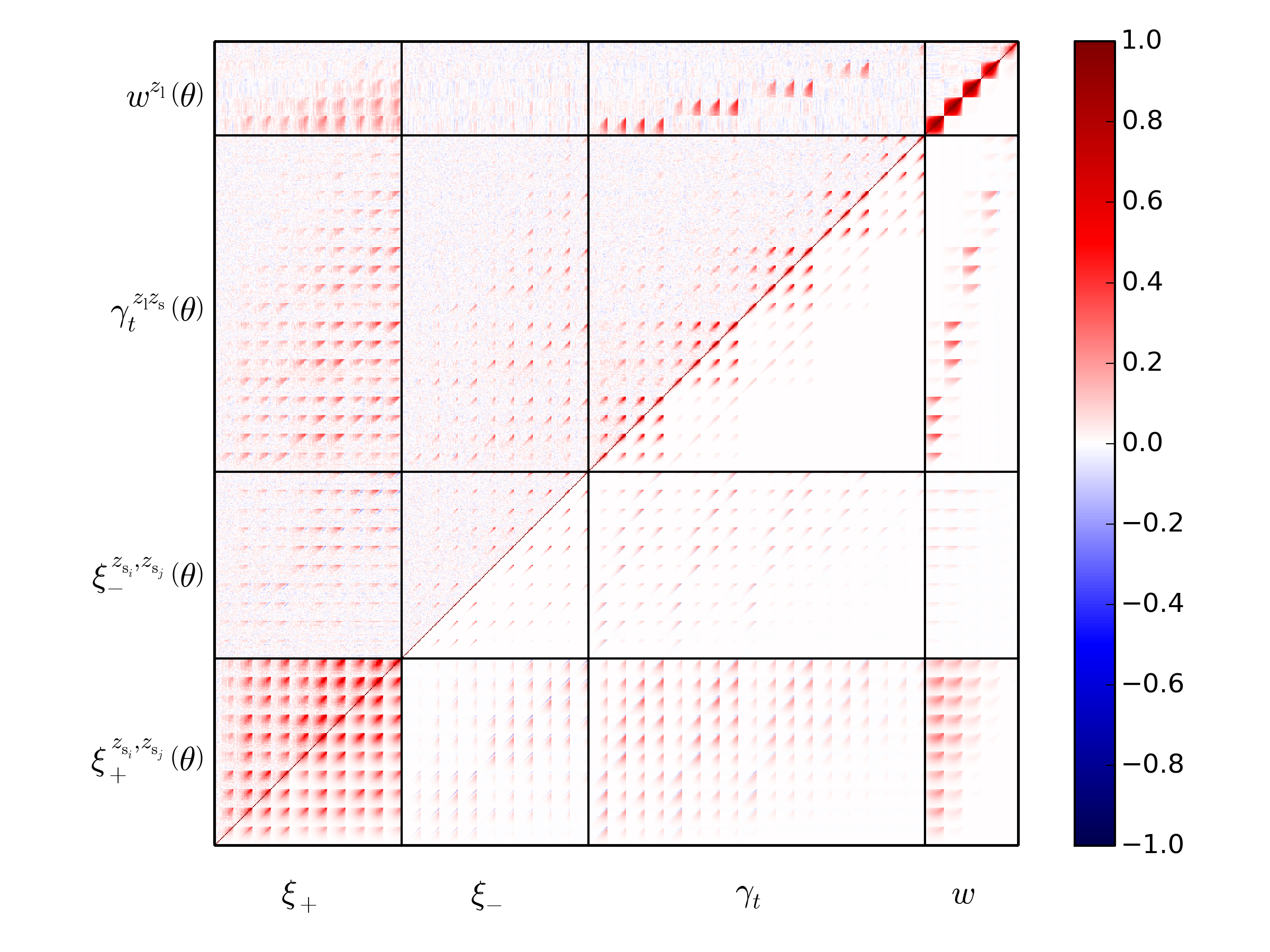}
  \end{minipage}\hfill
  \begin{minipage}{0.17\textwidth}
   \caption{Multi-probe correlation matrix for a joint data vector of cosmic shear, galaxy--galaxy lensing, and galaxy clustering including the non-Gaussian terms, with the same ordering as the data vector shown in Fig.~\ref{fig:cc1}. The upper left triangle shows the correlation matrix obtained from 1200 lognormal realizations (see Sect.~\ref{sec:covval} for details), the lower right shows the correlation matrix of the non-Gaussian halo model covariance (see Sect.~\ref{sec:covhm}). We recommend a zoom factor of $\sim$ 5 to inspect structures within the matrix.} \label{fi:covstruct}
  \end{minipage}
\end{figure*}

\paragraph*{Estimation from simulations} Estimating the precision matrix from a set of large, high-resolution numerical simulations using a standard Maximum Likelihood estimator is computationally prohibitively expensive even for single probe analyses \citep{tjk13,dos13,taj14}; this is even more an issue for the multi-probe case, where covariances are substantially larger. The main reason for these computational costs is the intrinsic noise properties of the estimator, which means we require a large ensemble of independent realizations of numerical simulations. Promising approaches can be separated into two main categories. The first is data compression \citep[see e.g.][]{sek10,eks14, ahb17}, which reduces the dimensionality of the covariance matrix. Second, recently new estimators with significantly improved noise properties \citep[see e.g.][]{pos08, pwz16, joa17, fre17} are being explored. 

\paragraph*{Estimation from data} Estimating covariance matrices from the data directly (through bootstrap or Jackknife estimators) avoids any assumptions about cosmological or other model parameters that need to be specified in the numerical simulation approach (and in the theoretical modeling approach). However, given the limited survey area, it is difficult to obtain a sufficiently large number of regions of sky for the method to work, and it is unclear if these regions can be treated as independent. We refer to \cite{Norberg2009, fseg2016} for more details. 

\paragraph*{Analytical modeling/computation} The analytic computation of weak lensing covariances was detailed in \citet{svk02} and \citet{jse08}, which derive straightforward expressions for Fourier and configuration space covariances under the assumption that density field is Gaussian, so that the four-point correlation of the density field can be expressed as the product of two-point correlations. On small and intermediate scales this assumption is inaccurate; analytical expressions of non-Gaussian weak lensing covariances were derived in \citet{taj09} and \citet{sht09}. These expressions were generalized to a 3x2pt analysis in \citet{cosmolike}. The main advantage of an analytical (inverse) covariance matrix is the lack of a noisy estimation process, which substantially reduces the computational effort in creating a large number of survey realizations; the disadvantage is that the modeling of the non-Gaussian covariance terms, which employs a halo model \citep[e.g.,][]{coh01}, is less precise compared to sophisticated numerical simulations. 

For the Y1KP analysis we implemented the third option, analytical modeling, for several reasons. First, \textsc{CosmoLike}'s analytical covariance implementation is fast enough to compute a configuration space covariance with 810,000 elements in $\le$ 12h. Second, as noted above there is no estimator noise in this calculation, which among other advantages allows us to use Eq.~(\ref{eq:like}) instead of using a multivariate $t$-distribution \citep{seh16}. Third, the non-Gaussian terms in our covariance are sub-dominant and hence corresponding uncertainties are unimportant (c.f. Fig.~\ref{fi:covariance_tests}). In general, non-Gaussian terms substantially impact covariance matrices, however our analysis excludes small scales (c.f. Sect.~\ref{sec:cuts}), and, as a consequence of the relatively low number density of source and lens galaxies, the noise terms in the covariance are comparatively large. 

In the following we summarize the analytic covariance computation, and validate the covariance matrix using Gaussian and log-normal simulations. We note that in the actual Y1KP data analysis we will use an iterative approach in order to account for the unknown underlying model of the analytical covariance matrix. Following \cite{esh09} we will update our fiducial covariance parameter set (c.f. Table \ref{tab:params}) with the best-fit parameters of the initial likelihood analysis run, and then rerun the likelihood analysis to obtain our final results. This procedure does not fully account for the cosmology dependence of the covariance matrix \citep[see][]{esh09,mos13,jts13}, but given the relatively large noise terms, this effect is not significant for DES Y1. It will be more important for future DES analyses.

\subsection{Halo Model Covariances}
\label{sec:covhm}
The covariance of two angular two-point functions $\Xi,\Theta \in \left\{w,\gamma_\mathrm{t}, \xi_+, \xi_-\right\}$ is related to the covariance of the angular power spectra by
\begin{widetext}
\be
\mathrm{Cov}\left(\Xi^{ij}(\theta),\, \Theta^{km}(\theta')\right) = \int \frac{dl\,l}{2\pi} J_{n(\Xi)}(l \theta) \int \frac{dl'\,l'}{2\pi}  J_{n(\Theta)}(l' \theta')\left[\mathrm{Cov}^{\mathrm{G}}\left(C_{\Theta}^{ij}(l),C_{\Xi}^{km}(l')\right)+\mathrm{Cov}^{\mathrm{NG}}\left(C_{\Theta}^{ij}(l),C_{\Xi}^{km}(l')\right)\right]\,,
\label{eq:FT_cov}
\ee
\end{widetext}
with $C_{\xi_+} \equiv C_{\xi_-} \equiv C_{\kappa\kappa}$, $C_{\gamma_\mathrm{t}} \equiv C_{\delta_{\mathrm{g}}\kappa}$ and $C_{w} \equiv C_{\delta_{\mathrm{g}}\delta_{\mathrm{g}}}$ in the notation of Eqs.~(\ref{eq:Cell}), and where the order of the Bessel function is given by $n = 0$ for $\xi_{+}$, $w$, $n = 2$ for $\gamma_\mathrm{t}$, and $n = 4$ for $\xi_{-}$. We calculate the covariance of the angular power spectra $\mathrm{Cov}\left(C_{\Theta}^{ij}(l),C_{\Xi}^{km}(l')\right)$ as the sum on Gaussian $\mathrm{Cov}^{\mathrm{G}}$ and non-Gaussian covariance $\mathrm{Cov}^{\mathrm{NG}}$, which includes super-sample variance \citep{tah13}, as detailed in \citet{cosmolike}, using the halo model to compute the higher-order matter correlation functions. Equation~\ref{eq:FT_cov} gives the covariance of two-point functions at angles $\theta$ and $\theta'$, and does not account for the finite width of angular bins. In practice, the covariance of two-point functions in angular bins is often evaluated at representative angles for each bin, assuming that the covariance varies only slowly across angular bins (called the narrow-bin approximation). The harmonic transform of the Gaussian contribution in Eq.~(\ref{eq:FT_cov}) reduces to a single integral as different harmonic modes are uncorrelated in the Gaussian covariance approximation. In the evaluation of the Gaussian covariance we split off the pure white noise terms and transform these terms analytically \citep{jse08}.

\subsection{Covariance Validation}
\label{sec:covval}
Most analytic models for the covariance of two-point functions in configuration space are assume the narrow-bin approximation, and that the maximum angular scales are much smaller than the survey diameter \citep[e.g.][]{svk02, hhs11, ccg11}. In the context of harmonic space correlation functions the latter assumption is also referred to as the $f_{\mathrm{sky}}$-approximation \citep{rpccg11}. Furthermore, masking and a non-compact survey geometry can also change the effective area of a survey as opposed to the simplifying assumptions made in our covariance model \citep{kil04, tstk14, lla16}.

To test the impact of binning, masking and survey geometry on the covariance matrix, we compare our covariance model to the sample covariance derived from different simplified realizations of our data vector. To generate the latter we used the \verb|FLASK| simulation tool \citep{xaj16}, which produces correlated Gaussian and log-normal random fields mimicking the projected density contrast and the lensing convergence of tomographic redshift bins. We generate a set of 150 Gaussian and 150 log-normal all-sky realizations using a fiducial $\Lambda$CDM cosmology (for configuration details see \citep{gglpaper,shearcorr}, where the same simulations are used to perform a number of systematics tests for galaxy--galaxy lensing and cosmic shear). From each all-sky realization 8 areas each the size of the DES-Y1 footprint are cut out, leading to a total of 1200 mock realizations of the DES-Y1 footprint.

To assess the agreement between different covariances we choose the Fisher formalism. For a set of parameters $\mathbf p$ and a covariance matrix $\mathbf{C}$ we compute the Fisher matrix
\begin{equation}
\label{eq:Fisher_matrix}
\mathrm{F}_{ij} = \left(\frac{\partial \mathbf M\left(\mathbf p\right)}{\partial p_i}\right)_{\mathbf{p}_0}^T \mathbf{C}^{-1} \left(\frac{\partial \mathbf M\left(\mathbf p\right)}{\partial ip_j}\right)_{\mathbf p_0}\ ,
\end{equation}
where $\mathbf p_0$ is the fiducial cosmology used to generate the \verb|FLASK| simulations and where the inverse of any sample covariance estimate has to be de-biased using the Kaufman-Hartlap correction \citep{hss07, Kaufman}.

\begin{figure*}
  \includegraphics[width=\textwidth]{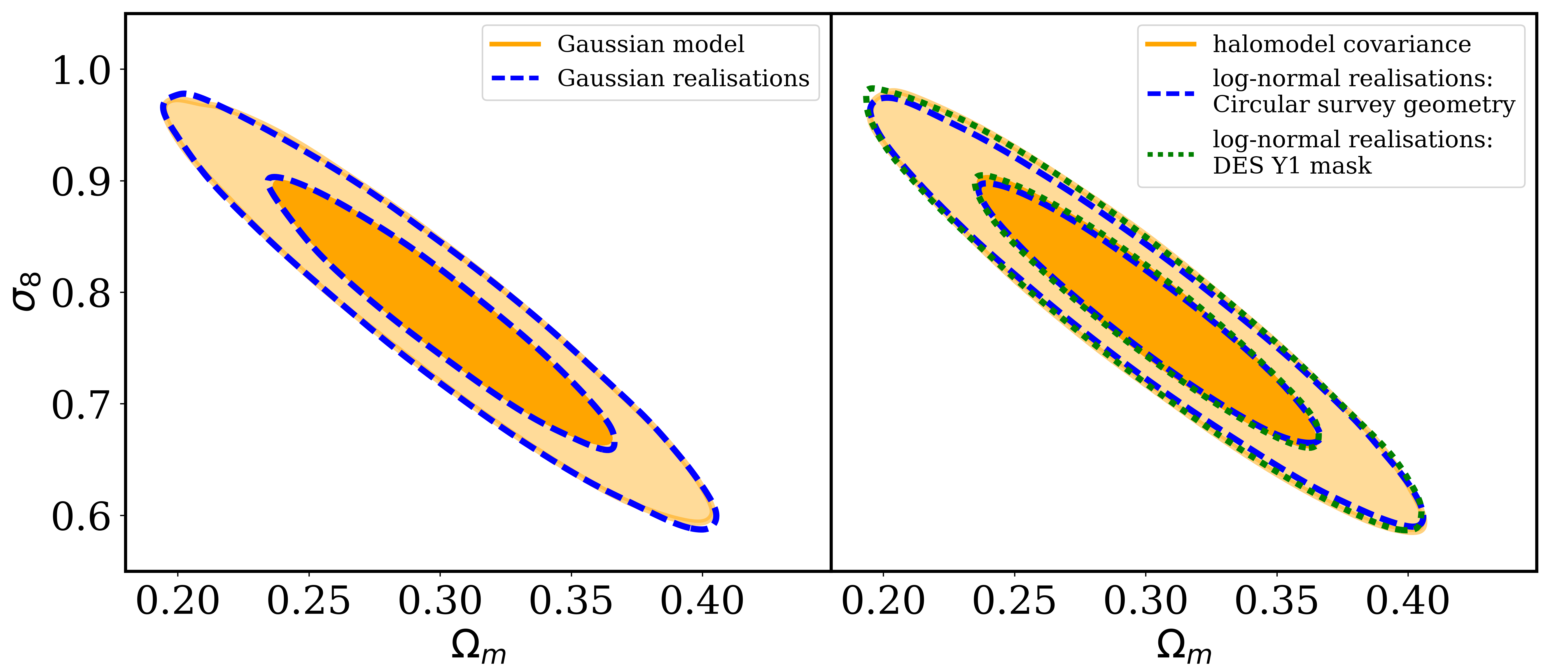}
 \caption{\emph{Left}: Comparison of 1$\sigma$ and 2$\sigma$ Fisher matrix-based parameter contours in the $\Omega_m $-$\sigma_8 $ plane derived from the Gaussian part of our covariance model (solid orange) to contours derived from Gaussian \textsc{FLASK} realizations (dashed blue line). These contours are marginalized over $24$ cosmology and nuisance parameters and include the parameter priors of the current version of the Y1KP analysis. \emph{Right}: Parameter contours derived from the sample covariance of two sets of differently shaped log-normal mock catalogs (dashed blue, dotted green lines), and contours obtained from the halo model covariance matrix (solid orange).}
\label{fi:covariance_tests}
\end{figure*}

Figure~\ref{fi:covstruct} provides a visual comparison of the correlation matrices obtained from log-normal realizations and the halo model. In Fig. \ref{fi:covariance_tests} we compare the Fisher contours on $\Omega_\mr m $ and $\sigma_8 $, the best-measured parameter combination, derived from different covariance matrices to test the above mentioned aspects of our covariance modeling. To derive these contours we marginalize over all 24 other $\Lambda$CDM cosmology and nuisance parameters. In particular, this marginalization also includes the priors of the current Y1KP data analysis.

On the left panel we show the contours derived from the Gaussian parts of our covariance modeling (orange, $\mathrm{Cov}^{\mathrm{G}}$ in Eq. \ref{eq:Fisher_matrix}) and the contours derived from the sample covariance of Gaussian \verb|FLASK| simulations. Both contours agree very well, which is a strong validation of our Gaussian covariance modeling. We find, however, that the narrow-bin approximation employed in Eq.~\ref{eq:FT_cov} overestimates the variance at large angular scales (by up to $50\%$ for the correlation functions $\xi_-(\theta)$ and $\gamma_t(\theta)$). We correct for this, as proposed in \citet[][Sect.~2.2.3]{fseg2016}, by computing our model covariance for a refined angular binning and then re-binning the resulting matrix to the actual angular binning.

The right panel of Fig.~\ref{fi:covariance_tests} compares Fisher contours derived from the sample covariance of two different sets of log-normal simulations to our complete halo-model covariance, again marginalizing over other cosmological and nuisance parameters and including the parameter priors of our final cosmological analysis. The first set of log-normal simulations is analyzed assuming a circular footprint that has the same area as DES-Y1 while for the second set of simulations we used the exact DES-Y1 footprint. The covariance of the Y1-shaped patches leads to marginally higher parameter uncertainties, but the disagreement is negligible compared to our overall constraining power. Furthermore, the contours derived from the halo model covariance agree with the contours from the log-normal covariances. 

Finally, we also tested our complete likelihood pipeline using cosmological N-body simulations, populated with realistic galaxy populations and designed to mimic the DES Y1 sample. This validation is described in a companion paper, \cite{simspaper}.

\section{Analysis Choices}
\label{sec:choices}
In this section we examine the robustness of the systematics modeling assumptions of the baseline model outlined in Sect.~\ref{sec:sys}. We stress that accurate treatment of systematic effects given the DES-Y1 statistical constraining power is challenging. In order to avoid parameter biases in the cosmology analysis we pursue two separate systematics mitigation strategies. First, we determine angular scale cuts that minimize the impact of known, but unaccounted-for systematic uncertainties in Sect.~\ref{sec:cuts}. Second, we mitigate systematic effects through marginalization over nuisance parameters (c.f. Sect.~\ref{sec:sys}). We stress-test both aspects of our systematics mitigation strategy in this section.

\subsection{Angular Scale Cuts}
\label{sec:cuts}
\begin{figure}
  \includegraphics[width=8.5cm]{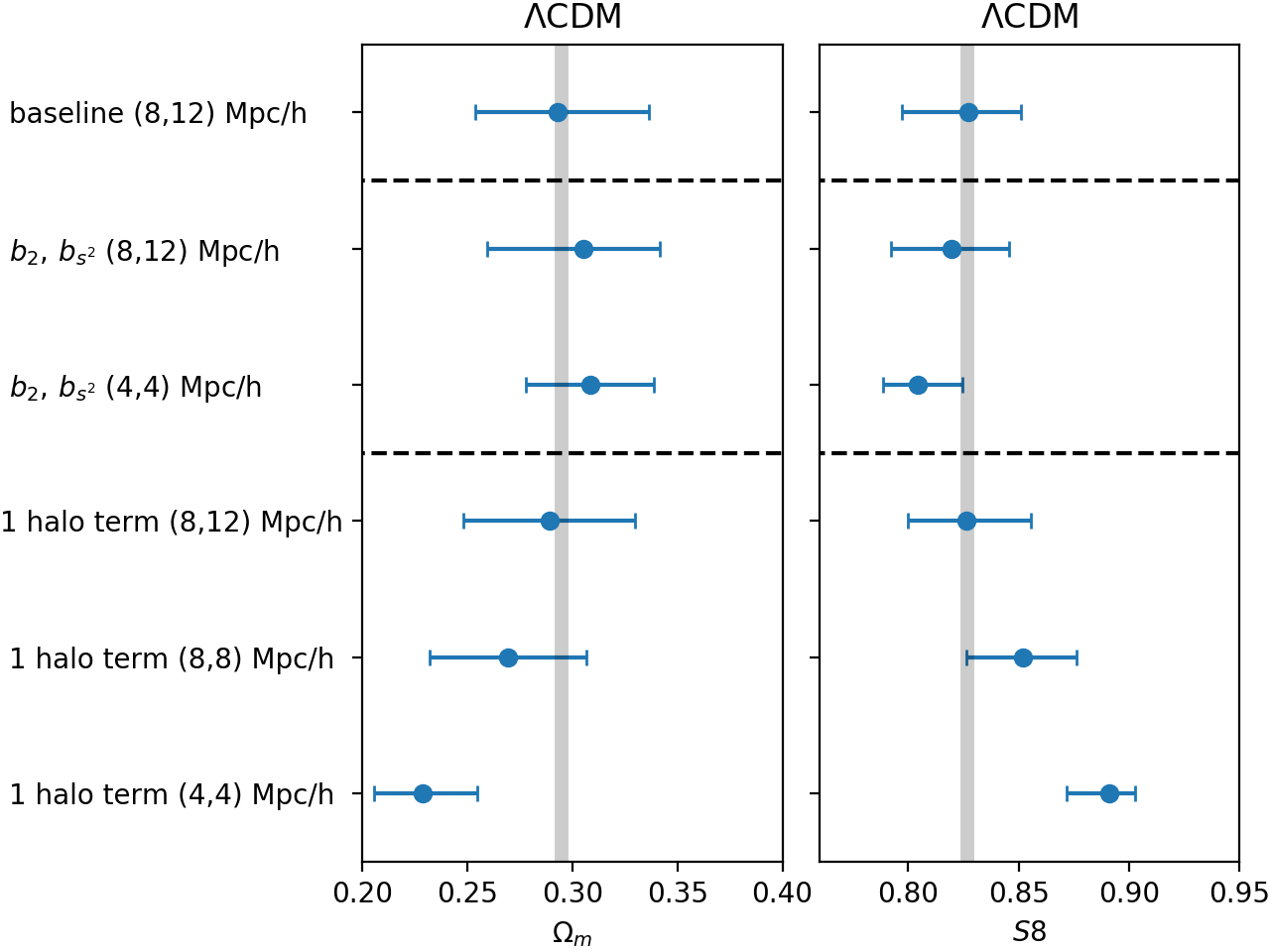}
\caption{Bias in cosmological parameters $\Omega_\mathrm{m}$ (left) and $S_8$ (right) due to unaccounted-for non-linearities in the data vector, for different scale cuts  $(R_\mathrm{clustering},R_\mathrm{ggl})$. The vertical grey line indicates the cosmology of the input data vectors; data points and error bars show the inferred cosmology parameters and $1\sigma$ uncertainties. The top line shows the constraining power of the baseline analysis, and demonstrates that it is unbiased. The next two lines show the parameter bias due to unaccounted-for non-linear galaxy biasing, and the bottom three lines show the parameter bias due to unaccounted-for contributions from the 1-halo term to $\gamma_\mathrm{t}$. See Sec.~\ref{sec:cuts} for details.}
 \label{fig:scalecuts}
\end{figure}
On small scales, accurate modeling of non-linearities of the density and galaxy fields is the key limitation of our baseline model. We seek to determine a set of scale cuts such that non-linear modeling limitations do not bias the cosmology results. As described in detail in Sect.~VIIIA of \citet{shearcorr}, the small-scale cuts on the shear correlation functions are determined to avoid parameter biases in weak lensing cosmology analysis due to baryonic feedback effects on the matter power spectrum. The baseline matter power spectrum model does not account for these effects and we defer a corresponding extension of the baseline model to future work (but see e.g. \citep{eifler15, mead15, maccrann17} for mititgation strategies).

The 3x2pt analysis adopts the same scale cuts on the shear correlation functions as the weak lensing analysis; this section focuses on scale cuts for the galaxy clustering and galaxy--galaxy lensing parts of the data vector. We define scale cuts in terms of a specific comoving scale $R$, and calculate the angular scale cut $\theta_{\mathrm{min}}^i$ for lens tomographic bin $i$ as
\be 
\theta_{\mathrm{min}}^i = \frac{R}{\chi\left(\langle z^i\rangle\right)}\,,
\ee
with $\langle z^i\rangle$ the mean redshift of galaxies in redshift bin $i$. Non-linear effects may impact galaxy clustering and galaxy--galaxy lensing differently, and we introduce separate scale cuts $R_\mathrm{clustering}$ and $R_\mathrm{ggl}$, which we report in the order $(R_\mathrm{clustering},R_\mathrm{ggl})$.

We determine conservative scale cuts using the following numerical experiments: 
\begin{enumerate}[label={\arabic{enumi}.}]
\item Generate data vectors that include additional non-linearities,
\item Analyze these data vectors with the baseline 3x2pt pipeline (that does not include these non-linear effects in the theoretical model),
\item Measure the bias in cosmology parameters due to unaccounted-for non-linearities,
\item Repeat 2, 3 for different scale cuts .
\end{enumerate} 

Figure~\ref{fig:scalecuts} summarizes the bias in cosmology parameters as function of scale cuts for two types of non-linearity:
\begin{enumerate}
\item Non-linear galaxy bias: We generate an input data vector that includes the next-to-leading order contributions to galaxy clustering and galaxy--galaxy lensing from quadratic bias $b_2$ and tidal bias $b_{\mathrm{s}^2}$ \citep{mcr09,Baldauf12}, which are evaluated using the \textsc{FAST-PT} code \citep{fastpt}. This data vector is then analyzed using the baseline model 3x2pt pipeline (assuming linear galaxy bias only).
\item Non-locality of $\gamma_\mathrm{t}$: Tangential shear is non-local, and contributions from deeply non-linear regime to $\gamma_\mathrm{t}$ are significant far beyond the halo radius (c.f. \citep{Baldauf10} for a detailed discussion). The contribution from an enclosed mass distribution of mass $M$ falls off as $M/R^{2}$, and we generate an input data vector that includes the 1-halo term contribution to $\gamma_\mathrm{t}$ based on the mean halo mass $\langle M_\mathrm{h}^i\rangle$ of the lens sample in redshift bin $i$ determined from realistic DES mock catalogs \citep{buzzard}, $\langle M_\mathrm{h}^i\rangle=\{3.23,\, 3.04,\,2.85,\,2.71,\,2.54\}\times 10^{13}M_\odot~h^{-1}$. We note that not all host halos are resolved, so these estimates provide an upper limit to the non-local contamination of $\gamma_\mathrm{t}$.
\end{enumerate}
Based on the data points in Fig.~\ref{fig:scalecuts}, we adopt a scale cut of 
\be
\left(R_\mathrm{clustering},R_\mathrm{ggl}\right)=(8,12)\,\mathrm{Mpc}~h^{-1}
\ee
 to avoid parameter biases due to non-linear biasing or non-locality of $\gamma_\mathrm{t}$.

\subsection{Stress-testing the baseline model}
\label{sec:sys2}
\begin{figure*}
  \includegraphics[width=17cm]{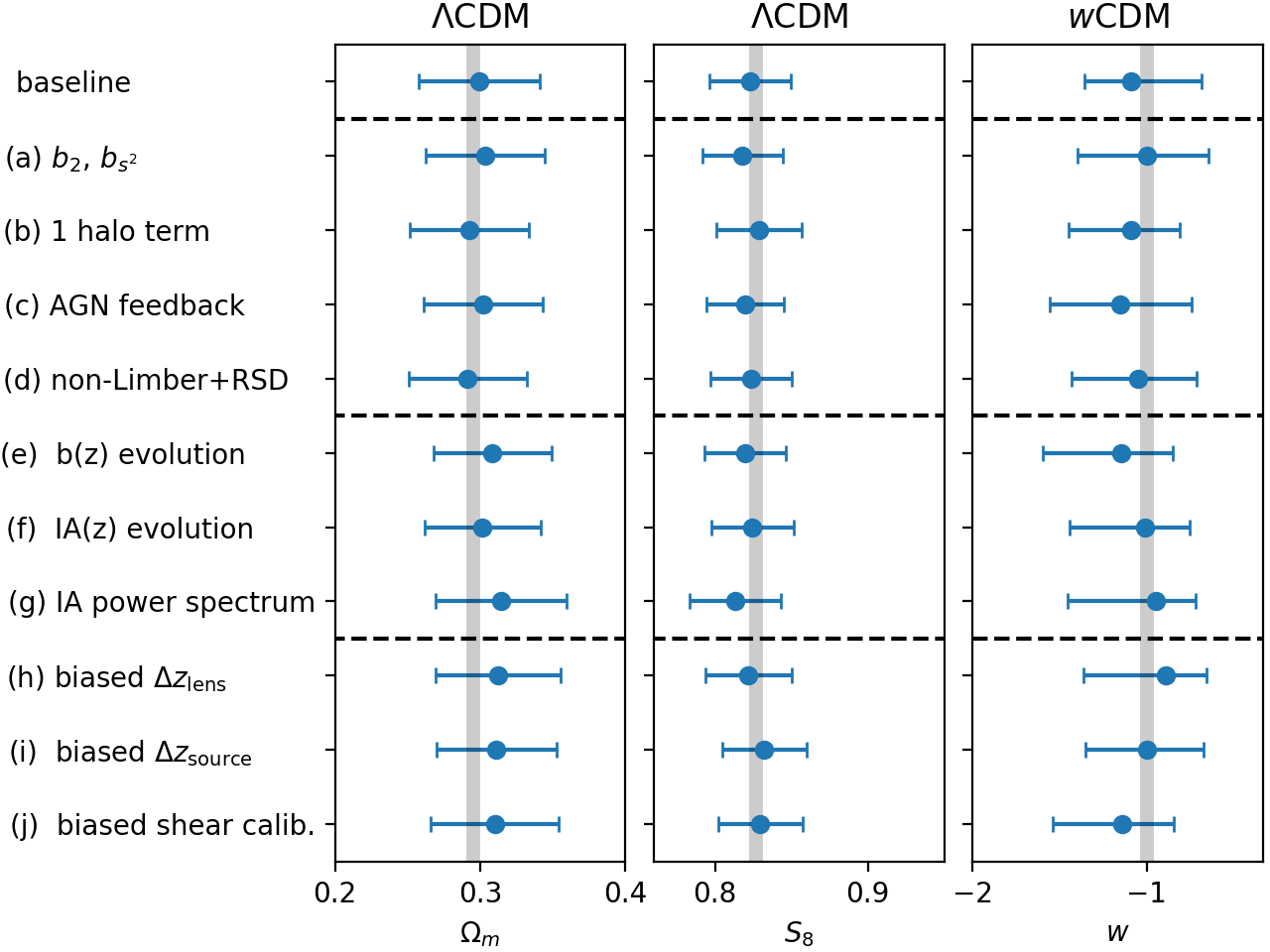}
\caption{Bias in cosmological parameters $\Omega_\mathrm{m}$ (left), $S_8$ (middle), and $w$ (right) due to different systematic uncertainties. The vertical grey line indicates the cosmology of the input data vectors; data points and error bars show the inferred cosmology parameters and $1\sigma$ uncertainties. The top line shows the baseline analysis for reference, the other scenarios are described in Sect.~\ref{sec:sys2}. Each source of error results in an impact on the parameters shown that is less than 0.5 $\sigma$}
 \label{fig:sys}
\end{figure*}
We now analyze the approximations and modeling choices of the baseline model using the same numerical technique as in the previous section. Figure~\ref{fig:sys} quantifies the parameter bias due to three groups of systematic biases --- physical effects not included in the baseline model, choice of parameterizations for systematic effects adopted in the baseline model, and the effect of misestimated priors on systematic effects --- which we discuss in turn.

The data points in the first four lines illustrate the parameter bias from known physical effects that are not included in the baseline model:
\begin{enumerate}[label=({\alph*})]
\item Non-linear galaxy bias: repeated from Sec.~\ref{sec:cuts} for completeness.
\item Non-locality of $\gamma_\mathrm{t}$: repeated from Sec.~\ref{sec:cuts} for completeness.
\item Baryonic feedback effects on the matter power spectrum: the input data vector is based on matter power spectrum from the \emph{AGN} scenario of the OWLS \citep{sdb10} suite of cosmological, hydrodynamical simulations, which includes baryonic feedback from supernovae, and AGN, and analyzed with the \textsc{halofit} baseline model. We stress that we assume all probes in our data vector to be affected by AGN feedback, including the galaxy-galaxy lensing and galaxy clustering part. This is a  conservative approximation, as redMaGiC galaxies form early and are likely less affected by feedback processes. We also note that the AGN scenario is considered to be one of the most extreme baryonic physics scenarios. Hence, the resulting bias seen in Fig. ~\ref{fig:sys} is an upper limit of baryonic effects. 
\item Limber approximation: We calculate the input data vector using the exact (non-Limber) expression, including the redshift-space distortion contributions to galaxy clustering \citep{Padmanabhan07}, and analyze it with the baseline model which employs the Limber approximation.
\end{enumerate}

The baseline models for astrophysical systematics (galaxy bias, intrinsic alignments) are somewhat arbitrary choices, and we now test whether these parameterizations are flexible enough to mitigate plausible variations of these models:

\begin{enumerate}[label=({\alph*})]
\setcounter{enumi}{4} 
\item Redshift evolution of linear galaxy bias: In addition to the scale dependence of galaxy bias discussed in the previous subsection, the redshift evolution of galaxy bias is another key uncertainty. Various fitting functions and physically motivated parameterizations for the redshift evolution of linear bias exist in the literature (see \citep{Clerkin15} for an overview); choosing among these is highly specific to the galaxy sample. DES-SV observations \citep{kwan16} and DES mock catalogs \citep{buzzard} 
indicate that the halo occupation distribution of the redMaGiC high-density sample evolves only weakly over the redshift range $0.1 < z < 0.6$. Hence the bias evolution of this sample is primarily caused by the redshift evolution of halo bias. We generate an input data vector that includes bias evolution within each redshift bin
\be
b^i(k,z) = b_1^i \times \frac{1+z}{1+\langle z^i\rangle}\,,
\ee
which is analyzed with the baseline model assuming constant bias within each redshift bin.
\item Redshift evolution of the IA amplitude: The NLA model is typically used to describe IA of early type galaxies (see e.g. \citep{tri14,Joachimi15} for recent reviews), often ignoring the alignment of blue galaxies, which is likely weaker \citep{hmi07,mbb11}. The observed IA of low-redshift, bright, red galaxies $A_{\mathrm{red}}(L,z)$ has been described as a power law in galaxy luminosity and redshift \citep{jma11,smm14}, although the theoretical expectation for redshift evolution is uncertain. To generate an expected IA amplitude redshift evolution (see \citep{keb16} for the detailed procedure), we calculate the mean IA amplitude of the red source galaxies $ \langle A_{\mathrm{red}}(m_{\mathrm{lim}},z)\rangle$ by averaging the observed amplitude scaling of \citet{jma11} over the DEEP2 red galaxy luminosity function \citep{fww07}, assuming a limiting magnitude $m_r \sim 23$ for the Y1 source sample. We then calculate an intrinsic alignment amplitude for the full source sample assuming no intrinsic alignments of blue galaxies, $A(z) =  \langle A_{\mathrm{red}}(m_\mathrm{lim},z)  \rangle \times f_{\mathrm{red}}(z)$ with $f_{\mathrm{red}}(z)$ the fraction of red galaxies, which is also estimated from the DEEP2 luminosity functions for red and all galaxies. With the observed IA normalization of \citet{jma11}, this IA contamination corresponds to $A_{\mathrm{IA},0} \sim 0.5$ at the pivot redshift $z_0$. Due to the rapid decrease of $f_\mathrm{red}$ with redshift, the resulting $A(z)$ is not monotonic in redshift. We generate an input data vector based on this $A(z)$, which is then analyzed with the baseline $A(z) \propto (1+z)^\alpha_{\mathrm{IA}}$ model.
\item IA power spectrum shape: Blue galaxies may align through tidal torquing \citep{ckb01}. These alignments are quadratic in the tidal field, and their power spectrum shape differs from the NLA model. We generate an input data vector contaminated with these quadratic alignments, which is then analyzed with the baseline NLA model. The amplitude of the quadratic alignment is chosen to give approximately the same IA contamination amplitude at 10 $\mathrm{arcmin}$ as the fiducial NLA model with $A_{\rm IA,0} = 1$, although the accuracy of this match depends strongly on scale and redshift bins. The IA modeling is described in more detail in \citep{btm17,shearcorr}.

\end{enumerate}

Finally, the last three lines in Fig.~\ref{fig:sys} show the impact of mis-estimating the mean of the Gaussian systematics priors by $1\sigma$ \emph{in each redshift bin}, on lens redshift shifts (h) source redshift shifts (i), and on shear calibration (j). Since we assume no correlation of these systematics across redshift bins, such a correlated shift corresponds to a several $\sigma$ misestimate of the prior.

We find that even the most agressive scenarios considered in this section bias the cosmology results by less that 0.5$\sigma$, and conclude that the baseline model with the scale cuts derived in Sect.~\ref{sec:cuts} is sufficiently flexible for the Y1KP analysis. We stress that this statement is based on the constraining power of the Y1KP analysis, and more detailed modeling will be required for future analyses.
\section{Sampler comparison}
\label{sec:samplers}

\begin{figure}
  \includegraphics[width=0.45\textwidth]{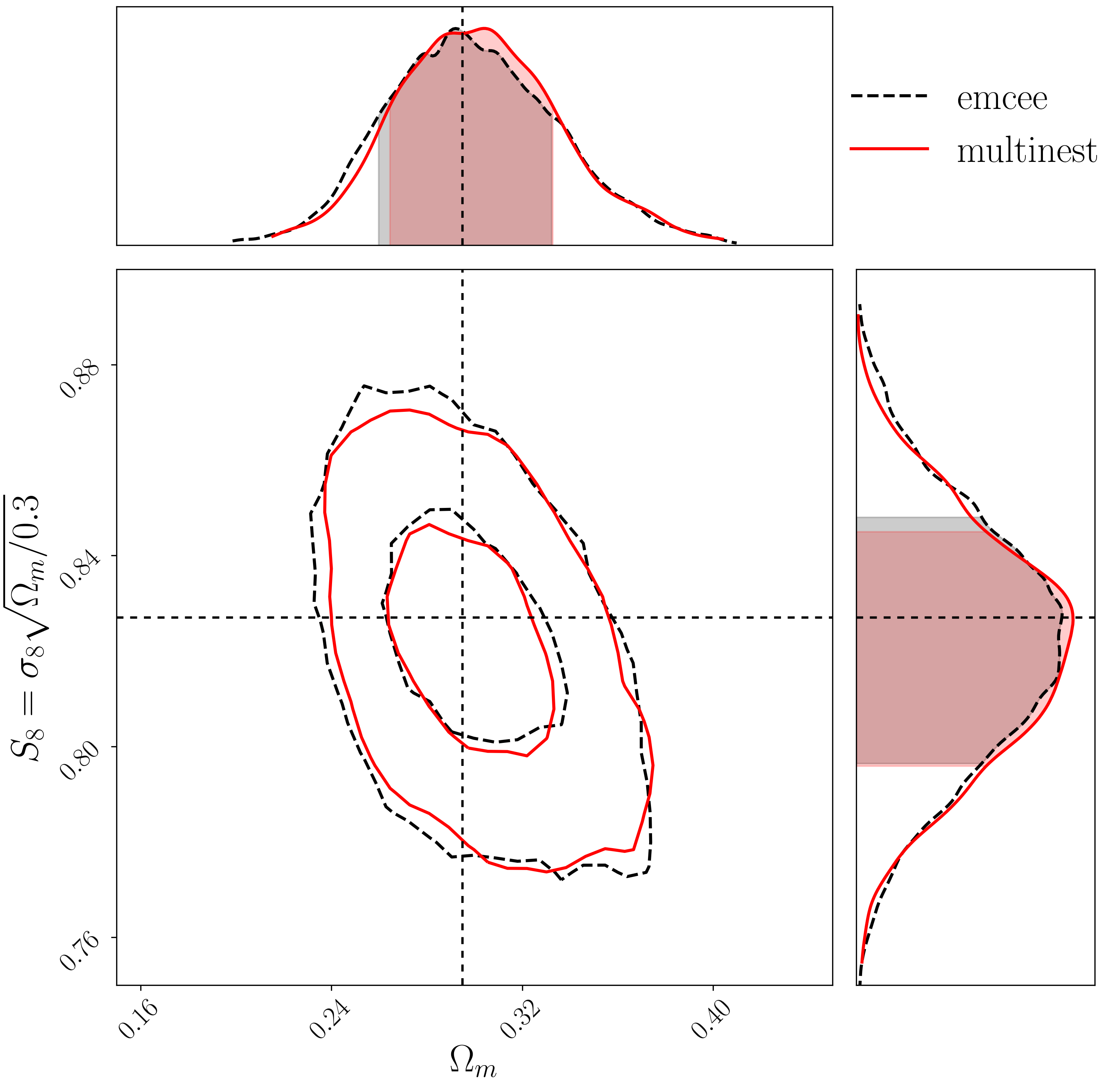}
  \caption{Comparison between parameter constraints obtained using two independent samplers, {\sc emcee} (black dashed contours) and {\sc multinest} (red solid contours).}
 \label{fig:sc}
\end{figure}
With validated model data vector and covariance at hand, we are now ready for parameter inference, the last step of the multi-probe analysis. Due to the high dimensionality of our parameter space, we adopt a sampling approach. We sample Eq.~(\ref{eq:like}) using two different samplers:

\paragraph*{{\sc Emcee}} \citep{foremanmackey,emcee}, uses the affine-invariant sampler of \citet{goodmanweare}, and can be parallelized with either MPI or shared memory multiprocessing \footnote{http://dan.iel.fm/emcee/current/}. To test its sampling convergence, we compare chains of lengths between $200,000$ and $2,000,000$ samples, and removed the burn-in phase (of typically $\sim100,000$ samples). Even after a chain is nominally converged, the parameter covariance and especially the $2\sigma$ contours may still evolve. We find that parameter contours typically stabilize after $~300,000$ samples, and the analyses presented in this paper use chains with $500,000$ samples. We have varied the number of \textit{walkers} and \textit{starting points} (including their variance) and found our results independent of reasonable choices in these settings.  

\paragraph*{ {\sc Multinest}}\citep{feroz}, which uses nested sampling \citep{skilling}, has a large number of tunable parameters which can strongly affect convergence rates. It is also designed to calculate Bayesian evidences, rather than just generating samples, like \emcee\ (though see \citealt{heavens17} for a method for extracting evidences from Monte Carlo Markov Chains in dimensions lower than the ones considered here). We use importance-nested sampling and a mono-modal likelihood in all runs, and vary four parameters for convergence studies: the number of live points $N_\mathrm{live}$, which controls the number of points in the ensemble, the efficiency \emph{eff}, which determines the rate at which the size of the sampling ellipse is decreased, the tolerance \emph{tol}, which determines the target evidence accuracy, and whether or not constant efficiency mode (\emph{const}) is enabled. The convergence is determined primarily by the tolerance parameter, and a covariance matrix error  corresponding approximately to 5\% accuracy in the posterior width, can be obtained with tolerance of $0.1$ and $>300$ live points. In this configuration we obtain about 1800 effective independent samples; if more are required, the number of live points can be increased.  With suitable configuration, we find that the parameter uncertainties obtained from both samplers agree at the few percent level (Fig~\ref{fig:sc}), comparable to the variance across chains; uncertainties due to the choice of sampling algorithm are negligible in the Y1KP error budget.

\section{Non-Mimimum Mass Neutrinos}
\label{sec:neutrinos}
Measurements of neutrino oscillations have established that neutrinos have mass, and provide a lower limit on the sum of the neutrino masses of $\Sigma m_\nu \gtrsim 0.06\,\mathrm{eV}$ \citep[see][for a review]{PDG16}. Neutrino mass affects the expansion history and distribution of mass and galaxies in the Universe, and cosmological observations provide tightest upper limits on the sum of neutrino masses: the Planck collaboration \citep{planck15} find $\Sigma m_\nu <0.49\,\mathrm{eV}$ from Planck CMB alone, and \citet{mnu15} find $\Sigma m_\nu <0.12\,\mathrm{eV}$ (all 95\% CL) from BOSS Lyman-$\alpha$ plus Planck. 

As the sum of neutrino masses is not precisely known yet, the Y1KP analyses marginalize over $\Omega_\nu h^2$ with a flat prior $[0.0006,0.01]$, corresponding to a conservative upper mass limit of $\Sigma m_\nu <1.0\,\mathrm{eV}$. In contrast, the previous sections of this paper present simulated DES-Y1 analyses with neutrino mass fixed to a value slightly above the minimum mass $\Sigma m_\nu = 0.06\,\mathrm{eV}$. We now consider the effects of this marginalization. 
For the DES-Y1 3x2pt data vector the dominant effect of massive neutrinos is the suppression of structure growth on small spatial scales. As described in Sect.~\ref{sec:code}, we implement this effect on the matter power spectrum using the \textsc{CAMB} and \textsc{CLASS} Boltzmann codes, which include the \citet{Bird12} fitting function for the impact of massive neutrinos on the non-linear matter power spectrum.

\begin{figure}
  \includegraphics[width=8.5cm]{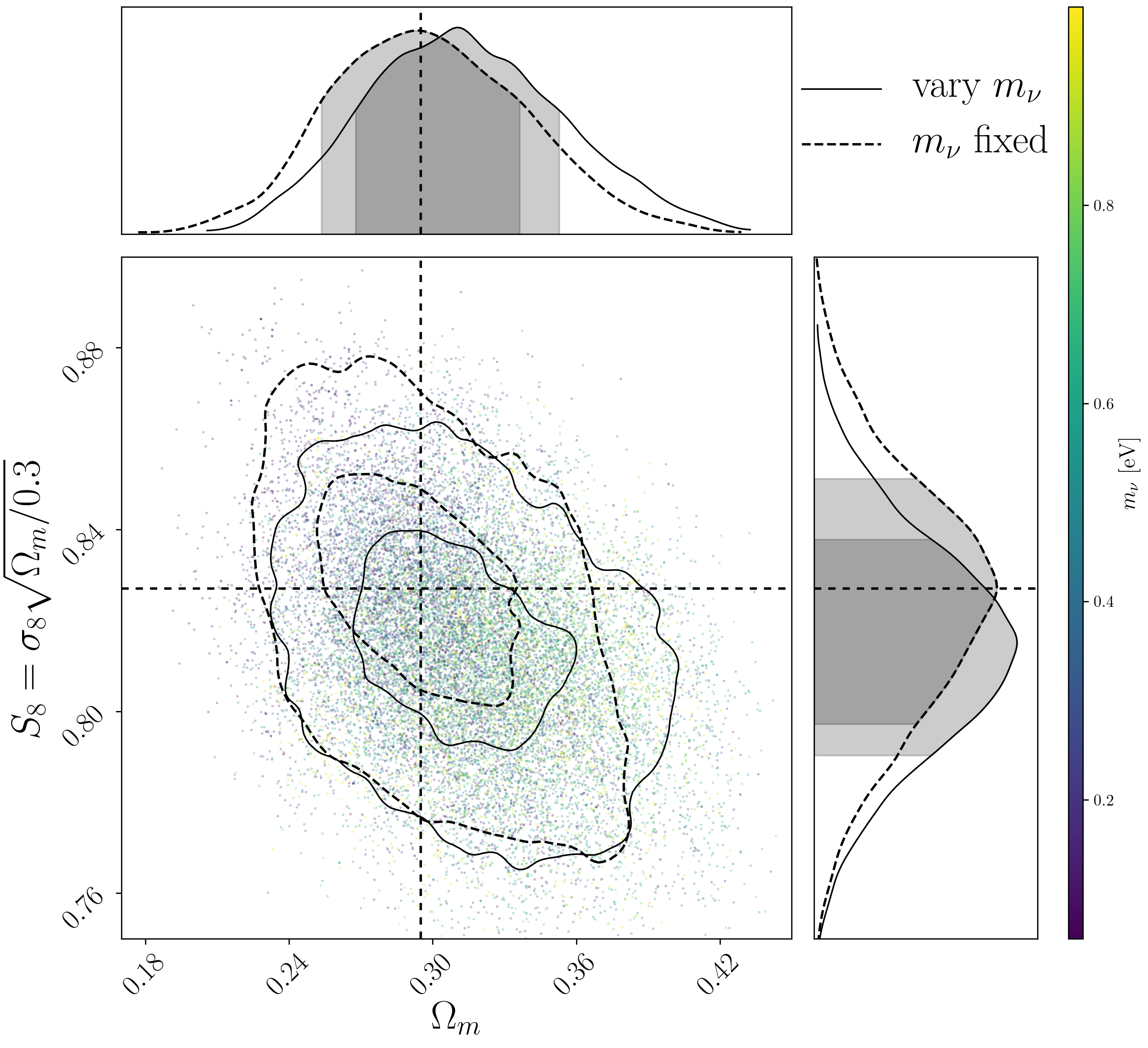}
  \caption{Parameter constraints for a simulated data vector at the fiducial cosmology on $\Omega_\mathrm{m}$ and $S_8$ with neutrino mass fixed at its minimum allowed value (dashed line), and marginalized over, assuming $\Omega_\nu h^2$ with the flat prior $[0.0006,0.01]$ (solid).}
 \label{fig:mnu}
\end{figure}

Figure~\ref{fig:mnu} shows parameter constraints for a simulated data vector at the fiducial cosmology (with minimum neutrino mass) on $\Omega_\mathrm{m}$ and $S_8$ with neutrino mass fixed (dashed line), and marginalized over $\Omega_\nu h^2$ (solid). The suppression of structure growth on small scales from massive neutrinos is degenerate with $\sigma_8$. As the prior on $\Omega_\nu h^2$ is not symmetric around its fiducial value, marginalizing over neutrino mass leads to shifts in the mean of the inferred parameter constraints \citep[see also][]{planck15,calabrese17}. The magnitude of this apparent bias in the marginalized parameters due to parameter degeneracies is scale dependent, which led us to fix neutrino mass in the baseline model in order to characterize parameter biases from unaccounted-for systematic effects in a setting where the fiducial analysis recovers the input parameters unbiased.

In addition to their effect on expansion history and matter power spectrum, massive neutrinos also introduce a scale dependence to halo bias \citep{bnu0,bnu1,bnu2}. The corresponding modification of galaxy bias
\be
b^i(k,z) \rightarrow b_\nu^i(k,z)
\ee
is implemented in the \textsc{CosmoSIS} pipeline using the analytic expressions of \citep{bnu2}. For the DES-Y1 3x2pt data vector with the $(8,12)\,\mathrm{Mpc}~h^{-1}$ scale cuts, this scale-dependent bias shifts the smallest-scale clustering and galaxy--galaxy lensing data points by up to $2.5\%$, and for the full data vector the shift amounts to $\Delta \chi^2 =0.1$ (at $\Omega_\nu h^2 = 0.01$) compared to an implementation without scale-dependent bias from massive neutrinos; the impact on $\Omega_m$ and $S_8$ constraints is minimal.  

In summary, 1) the fiducial DES Y1KP analysis does not constrain the sum of neutrino masses, 2) the numerical experiment presented in Fig.~\ref{fig:mnu} illustrates that marginalizing parameter space extensions may affect not only the uncertainty, but also shift the maximum likelihood value of baseline parameters, and 3) it is important to compare parameter constraints from different analyses and/or experiments within the same parameter space. 

\section{Conclusions}
\label{sec:conc}
Gains in cosmological constraining power for cosmic surveys will come from two main directions: the reduction in statistical uncertainties due to the larger survey area and increased depth of future surveys, and the improved methodology in the data analysis as it relates to combining correlated observables and modeling and/or mitigating systematic effects.
Our companion papers on DES Y1 data will advance the state-of-the-art in the first category; this work advances the state-of-the-art in the second category. 

We present here the necessary ingredients to conduct a joint cosmic shear, galaxy--galaxy lensing, and galaxy clustering analysis in configuration space that accounts for all cross-correlations amongst these probes. We developed two independent modeling pipelines, based on the \textsc{CosmoSIS} and \textsc{CosmoLike} frameworks, that allow us to cross-validate our implementation. This comparison was critical in identifying code-specific sources of uncertainties, including integration precision, interpolation precision, interpretation of histograms.  

For the first time in the literature, we demonstrate the capability to compute an analytic 3x2pt non-Gaussian covariance matrix in configuration space, and we validate said covariance using numerical simulations. We show that the impact of the DES Y1 mask on the covariance is negligible for the Y1KP analysis and we also show that choosing a different input cosmology for the covariance has minimal impact. 

We carry out realistic simulations of DES Y1KP analyses that jointly sample the $\Lambda$CDM/wCDM cosmological parameter space and 20 nuisance parameters (see Table \ref{tab:params}) accounting for uncertainties in lens and source photo-$z$, shear calibration, intrinsic alignments, and galaxy bias. Given our minimum scale cuts of 8 Mpc$~h^{-1}$ for clustering and 12 Mpc$~h^{-1}$ for galaxy-galaxy lensing, we show that the Y1KP analysis is robust against modeling uncertainties in the non-linear regime, but also against potential biases in estimating the means of shear calibration and  photo-$z$ bias parameters. We further examine the impact of different samplers (\textsc{Emcee}~ and \textsc{Multinest}) on parameter constraints and describe settings for both methods that ensure unbiased constraints.   

In summary, we have developed a new and comprehensive capability to conduct a joint cosmic shear, galaxy--galaxy lensing, and galaxy clustering analysis in configuration space and we have validated that our inference pipelines meet the precision required for the DES Y1KP. 

Future extensions of these pipelines will include more sophisticated modeling of non-linear scales in the correlation functions, e.g. to include non-linear bias and Halo Occupation Distribution models, and to properly account for baryonic effects. Such developments will allow us to utilize additional information in the quasi-linear regime. An extension of the data vector to include additional probes from photometric DES data, for example troughs (underdensities) and galaxy clusters (overdensities) can provide an additional avenue to increase the cosmological information content using the same survey data.

It is important to note that the analyses presented here only span the $\Lambda$CDM and wCDM parameter space, and a large variety of interesting fundamental physics questions that extend this parameter space can also be tested with DES multi-probe analyses. Extending our pipelines to model additional science cases, for example, modified gravity or interacting dark matter scenarios, will be a focus of future work.

Finally, we emphasize the importance of sophisticated science analysis software development for future cosmological data analyses. The increased statistical power of future data sets and the increased complexity of future analyses with respect to probes included and physics/systematics modeled, will require a change in how the community collaborates and builds analysis software. Independent implementation, cross-validation, and simulated analyses will be critical to achieve credible constraints on our cosmological model and it will require us to better interface expertise in statistical methods, numerical simulations, and software development. The two independent pipelines and the tests and simulated analyses presented in this paper are a first step in this direction, but far from sufficient for future precision analyses.

\section*{Acknowledgments}

The figures in this work are produced with plotting routines from matplotlib \citep{matplotlib}, ChainConsumer\citep{chainconsumer}, and Emmanuel Schaan.

EK was supported by a Kavli Fellowship at Stanford University. Part of the research was carried out at the Jet Propulsion Laboratory, California Institute of Technology, under a contract with the National Aeronautics and Space Administration and is supported by NASA ROSES ATP 16-ATP16-0084 grant and by NASA ROSES 16-ADAP16-0116. OF was supported by SFB-Transregio 33 `The Dark Universe' of the Deutsche Forschungsgemeinschaft (DFG) and by the DFG Cluster of Excellence `Origin and Structure of the Universe'.
EK, TE, and RHW thank the Kavli Institute for Theoretical Physics, supported in part by the National Science Foundation under Grant No.\ NSF PHY-1125915, for hospitality while this work was completed.

Funding for the DES Projects has been provided by the U.S. Department of Energy, the U.S. National Science Foundation, the Ministry of Science and Education of Spain, 
the Science and Technology Facilities Council of the United Kingdom, the Higher Education Funding Council for England, the National Center for Supercomputing 
Applications at the University of Illinois at Urbana-Champaign, the Kavli Institute of Cosmological Physics at the University of Chicago, 
the Center for Cosmology and Astro-Particle Physics at the Ohio State University,
the Mitchell Institute for Fundamental Physics and Astronomy at Texas A\&M University, Financiadora de Estudos e Projetos, 
Funda{\c c}{\~a}o Carlos Chagas Filho de Amparo {\`a} Pesquisa do Estado do Rio de Janeiro, Conselho Nacional de Desenvolvimento Cient{\'i}fico e Tecnol{\'o}gico and 
the Minist{\'e}rio da Ci{\^e}ncia, Tecnologia e Inova{\c c}{\~a}o, the Deutsche Forschungsgemeinschaft and the Collaborating Institutions in the Dark Energy Survey. 

The Collaborating Institutions are Argonne National Laboratory, the University of California at Santa Cruz, the University of Cambridge, Centro de Investigaciones Energ{\'e}ticas, 
Medioambientales y Tecnol{\'o}gicas-Madrid, the University of Chicago, University College London, the DES-Brazil Consortium, the University of Edinburgh, 
the Eidgen{\"o}ssische Technische Hochschule (ETH) Z{\"u}rich, 
Fermi National Accelerator Laboratory, the University of Illinois at Urbana-Champaign, the Institut de Ci{\`e}ncies de l'Espai (IEEC/CSIC), 
the Institut de F{\'i}sica d'Altes Energies, Lawrence Berkeley National Laboratory, the Ludwig-Maximilians Universit{\"a}t M{\"u}nchen and the associated Excellence Cluster Universe, 
the University of Michigan, the National Optical Astronomy Observatory, the University of Nottingham, The Ohio State University, the University of Pennsylvania, the University of Portsmouth, 
SLAC National Accelerator Laboratory, Stanford University, the University of Sussex, Texas A\&M University, and the OzDES Membership Consortium.

The DES data management system is supported by the National Science Foundation under Grant Number AST-1138766.
The DES participants from Spanish institutions are partially supported by MINECO under grants AYA2015-71825, ESP2015-88861, FPA2015-68048, SEV-2012-0234, SEV-2012-0249, and MDM-2015-0509, some of which include ERDF funds from the European Union. IFAE is partially funded by the CERCA program of the Generalitat de Catalunya.

YO acknowledges funding from the Natural Sciences and Engineering Research Council of Canada, Canadian Institute for Advanced Research, and Canada Research Chairs program. N.K. acknowledges support from FAPESP through grant 2015/20863-9.
Some of the computations in this paper were made on the supercomputer Guillimin from McGill University, managed by Calcul Qu\'{e}bec and Compute Canada. The operation of this supercomputer is funded by the Canada Foundation for Innovation (CFI), the minist\`{e}re de l'\'{E}conomie, de la science et de l'innovation du Qu\'{e}bec (MESI) and the Fonds de recherche du Qu\'{e}bec - Nature et technologies (FRQ-NT).
Parts of this research is carried out as part of the Blue Waters sustained-petascale computing project, which is supported by the National Science Foundation (awards OCI-0725070 and ACI-1238993) and the state of Illinois. Blue Waters is a joint effort of the University of Illinois at Urbana-Champaign and its National Center for Supercomputing Applications.

\bibliography{methods_paper} 

\appendix

\section{Code comparison}
\label{sec:codecomp}
In addition to the code comparison at a fiducial point in parameter space (c.f. Fig. \ref{fig:cc1}) and the comparison of the all encompassing simulated likelihood analyses (c.f. Fig. \ref{fig:cc2}), we also map the response of \textsc{CosmoSIS} and \textsc{CosmoLike} with respect to the individual parameter dimensions that enter our analysis. 

\begin{figure*}
   \includegraphics[width=17cm]{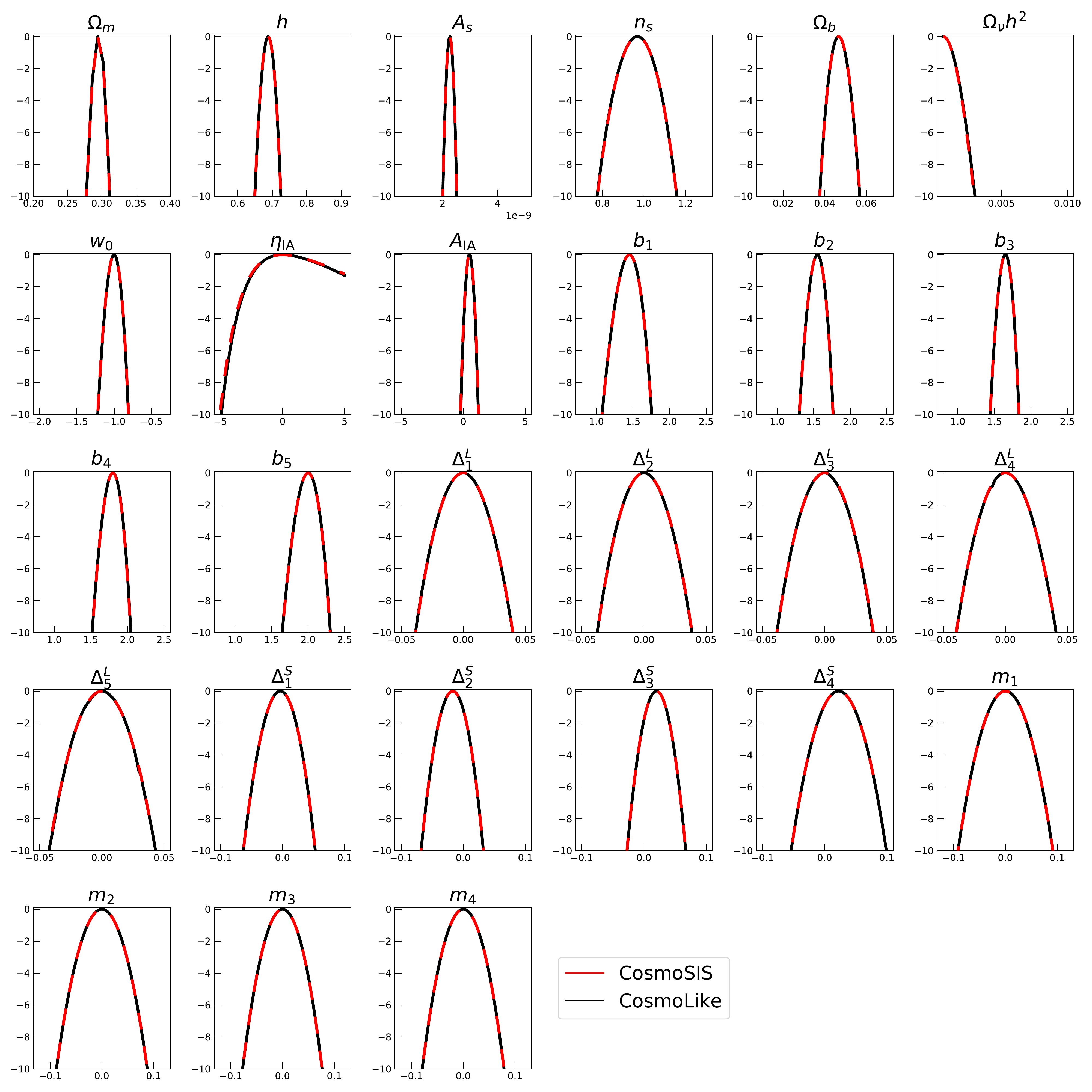}
 \caption{We show the response of the two different analysis codes in 27 dimensions, where in each analysis we fix 26 of the 27 parameters at their fiducial value. The y-axes of all panels show the log-likelihood as a function of the varied parameters. Red lines correspond to the \textsc{CosmoSIS} framework and black to \textsc{CosmoLike}.}
  \label{fig:coderesponse}
\end{figure*}

Figure \ref{fig:coderesponse} shows the log-likelihood of all 27 dimensions considered in our simulated analyses while fixing all other parameters to the fiducial values (see Table \ref{tab:params}). We find excellent agreement between the independently developed codes. We stress that this code comparison revealed several insufficiencies in both frameworks related to the interpretation of binned histograms,  precision in integration and interpolation routines, and subtle coding errors that would have likely remained hidden in the absence of a thorough testing scheme.

\end{document}